\documentclass{article}
\usepackage{epsfig}

\begin{document}
\null
\hfill DESY 02-111\\
\null
\hfill TUFTS CONF-02/21-1-E\\
\null
\hfill hep-ph/0210368\\
\vskip 1cm
\begin{center}
{\Large \bf Physics at large $p_T^2$ and $Q^2$:\\[.4em]
Summary\footnote{Plenary presentation at the X. International Workshop
on Deep Inelastic Scattering (DIS2002) Cracow, Poland, 30 April--4 May, 
2002.}  }\\[.4em]

\vskip 2.5em
{\large
{\sc 
G. Moortgat--Pick$^{a}$\footnote{
e-mail: gudrid@mail.desy.de}}, 
{\sc S. Rolli$^{b}$\footnote{e-mail: rolli@fnal.gov}}, 
{\sc
A.F. Zarnecki$^{c}$\footnote{
e-mail: Filip.Zarnecki@fuw.edu.pl}}
}\\
\vskip .5em
\end{center} 
\par
\vskip .1cm
\begin{center}
\begin{itemize}
\item[a)] II. Institute for Theoretical Physics,
University of Hamburg and\\ 
DESY, Deutsches Elektronen-Synchrotron, 22607 Hamburg, Germany
\item[b)] 
TUFTS University, Department of Physics \& Astronomy, \\
                        Medford, MA 02155, USA 
\item[c)] Inst. of Experimental Physics, Warsaw Univ., 
                        Ho\.za 69,\\ 00-681 Warszawa, Poland
\end{itemize}
\end{center}
\par
\vskip 1cm
\begin{abstract}
We summarize the results presented in the
{\it Physics at large $p_T^2$ and $Q^2$} working group 
at the DIS'2002 Workshop.
Higgs searches, precision measurements as well as searches for physics 
beyond the Standard Model  at current and future experiments
are reviewed.
\end{abstract}

\vspace{1em}

\vspace{1em}
\hfill

\newpage

\section{Introduction}

Over the past three decades the Standard Model (SM) of particle physics has been surprisingly
successful.
Although the precision of the experimental tests improved by orders
of magnitude no significant deviation from the SM predictions has
been observed besides the compelling evidence for non--zero neutrino masses.
Still, there are many questions which the Standard Model does not
answer and problems it can not solve.
Among the most important ones are the origin of the electro-weak
symmetry breaking, hierarchy of scales, unification of fundamental forces
and the nature of gravity.
Precise measurements of physics at highest $p_T^2$ and $Q^2$ values 
should finally help us to solve the puzzles of the Standard Model.
The discovery of the Higgs particle and the measurement of its properties
is the most important challenge of future experiments.
However, the signs of ``new physics'' can be also looked for in many
other channels.

In this paper we will report on the status of the experimental Standard Model tests,
including searches for top physics and searches for the SM Higgs boson.
Searches for 
supersymmetry, low-scale gravity, leptoquarks and other
new phenomena beyond the Standard Model, are reviewed.
Finally, the prospects for
discovering ``new physics'' at existing and future colliders
are discussed.


\section{Precision tests of the Standard Model}
\label{sec:sm}

Over the last decade the electroweak physics has entered the era of
precision measurements, resulting in experimental accuracies
better than the per mille level~\cite{saeki}: examples are the  W-boson mass,
measured at LEP and the TeVatron, and the effective weak mixing angle at
the Z~resonance, measured by SLD and at LEP.

The comparison of electroweak precision measurements with accurate
theory predictions allows to test the electroweak theory at the
quantum level, where all parameters of the model enter. In this way it
has been possible to obtain indirect constraints on the top-quark mass
prior to the top-quark discovery, which turned out to be in remarkable
agreement with the direct observation carried out at the
TeVatron. With the knowledge of the top-quark mass and further
improved experimental and theoretical precisions, it is now possible
to obtain constraints on the Higgs boson mass within the Standard
Model, which enters the precision observables in leading order only
logarithmically in contrast to the quadratic dependence on the top
quark mass.
 
\begin{figure}[h]
\begin{center}
  \begin{minipage}{2.3in}\hspace*{-.5cm}
\psfig{figure=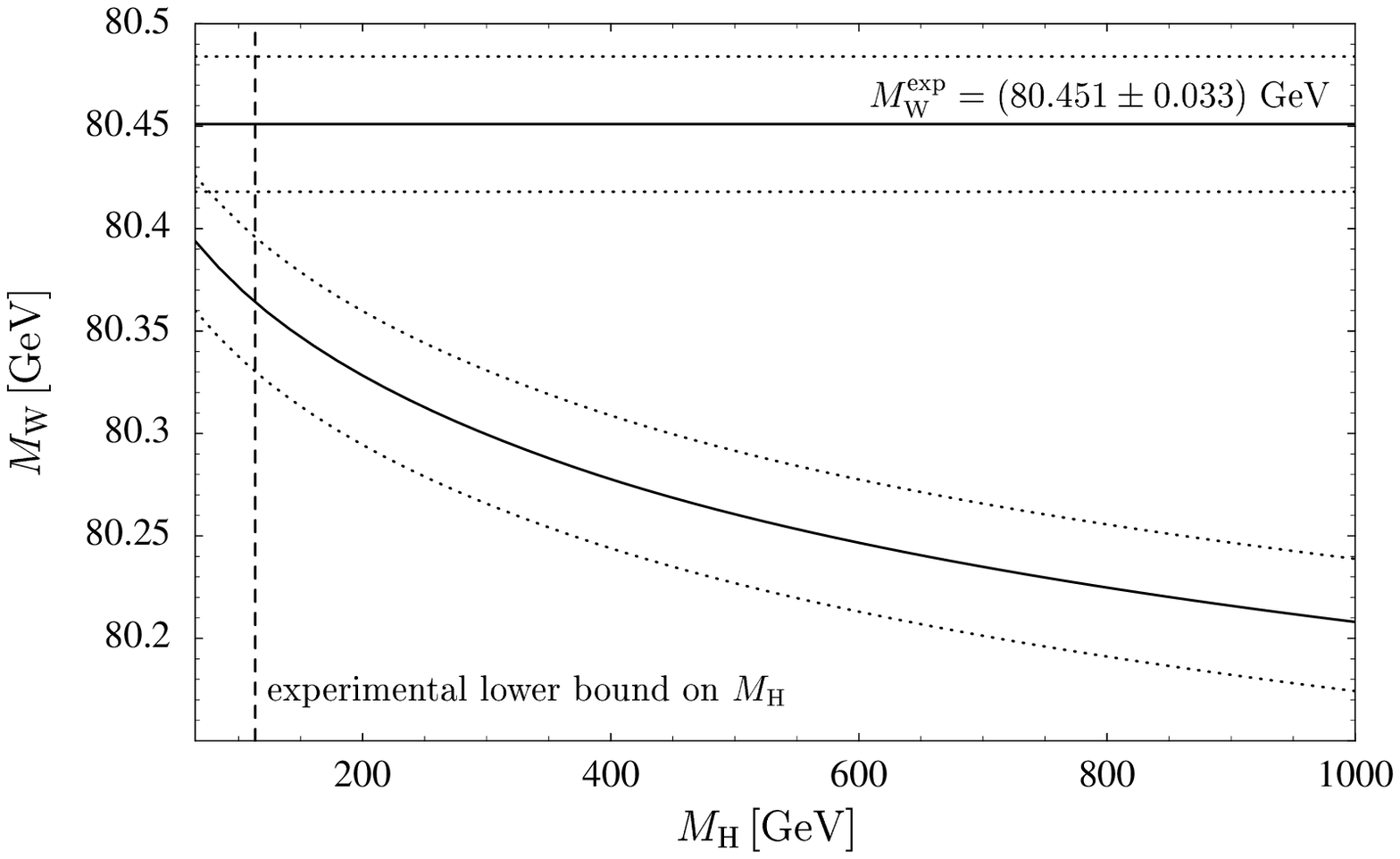,height=1.5in}
\caption{ {\scriptsize
Comparison of the theory prediction for $M_W$ as function of
the Higgs-boson mass with the current experimental value. The
experimental exclusion limit on the Higgs-boson mass is also indicated
(from Ref.~\cite{Weiglein}).}}
    \label{fig:mwtheoexp}
  \end{minipage}\vspace*{-.1cm}
\hfill
\hspace*{-.5cm}
  \begin{minipage}{2.3in}\hspace*{.5cm}
\psfig{figure=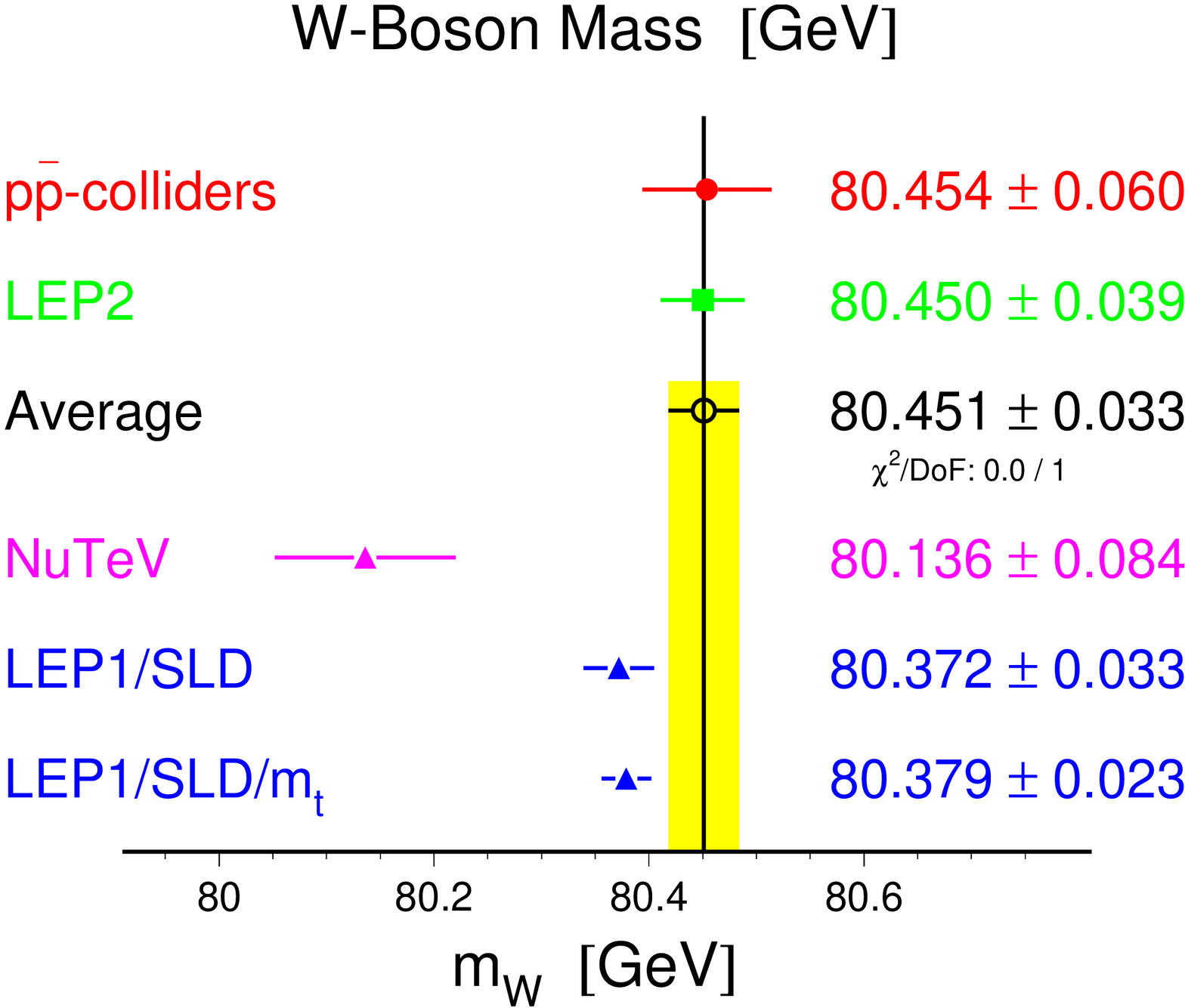,height=1.5in}
\caption{ {\scriptsize The world average of the direct $m_W$ 
           measurements from $p\bar{p}$ colliders and LEP2. 
           Also shown are indirect $m_W$ determinations within the
           Standard Model by NuTeV, LEP1+SLD and LEP1+SLD with 
           $m_t$ measurement.  }}
    \label{fig:comparison}
  \end{minipage}\vspace*{-.1cm}
\end{center}
\end{figure}

As an example for the comparison between theory and experiment,
Fig.~\ref{fig:mwtheoexp} shows the currently most accurate prediction
for the W-boson mass within the Standard Model~\cite{Weiglein}, 
derived from the
prediction for muon decay, in comparison with the current experimental
value for $M_W$ and the experimental exclusion limit on the Higgs-boson
mass, $M_H > 114.4$~GeV at 95\% C.L.. The theory uncertainty is dominated
by the experimental error of the top-quark mass, which enters the theory
prediction as input parameter, while the present uncertainty from unknown
higher-order corrections is smaller by about a factor 5~\cite{Weiglein}.
The figure clearly shows the preference for a light Higgs boson within
the Standard Model; at the $1 \sigma$ level there is no overlap between
the allowed regions of experimental result and theory prediction for
$M_H > 114.4$~GeV.

Fig.~\ref{fig:comparison} displays the experimental values from LEP
and the TeVatron and the world average obtained from combining these
two measurements, as well as indirect predictions from a Standard
Model fit to the LEP1+SLD data and the LEP1+SLD data supplemented by
the $m_t$ measurement. Furthermore shown is the prediction within the
Standard Model corresponding to the measurement from the NuTeV
collaboration, which has recently published its final result on the
ratio of neutral current to charged current reactions in
neutrino-nucleon scattering. This measurement, when interpreted as a
measurement of the mass of the W boson, shows an interesting
deviation, at the level of three standard deviations, from the direct
measurement.  The NuTeV experiment has extracted the electroweak
parameter, $\sin^2 \theta_W$, from the high precision measurement of
the ratio of neutral-current to charged-current cross sections in
deep-inelastic neutrino and anti-neutrino scattering off a steel
target.  Their measurement, $\sin^2 \theta_W^{on-shell} = 0.2277 \pm
00013(stat) \pm 0.0009(syst)$, is $3 \sigma$ above the Standard Model
prediction.  The plausibility of the hypothesis that this discrepancy
is due to unaccounted QCD effects, especially a strange and
anti-strange sea asymmetry has been evaluated by taking into account
results from NuTeV, CCFR, and charged-lepton deep-inelastic
cross section measurements. The NuTeV collaboration does not find
support for this hypothesis~\cite{panagiotis}.

\begin{figure}[h]
\begin{center}
\epsfig{figure=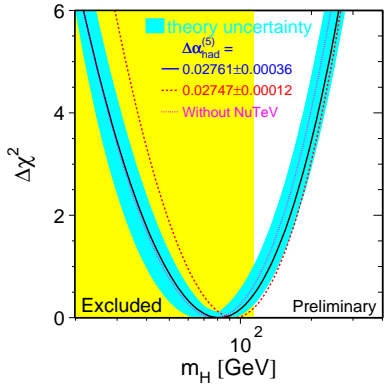,height=6cm,clip=}
\end{center}\vspace*{-1cm}
 \caption{Indirect Higgs mass constraints from a global fit to all data
in the SM.}
  \label{blueband}
\end{figure}

The current result of the global fit to all data in the Standard Model is shown
in Fig.~\ref{blueband}, where the $\Delta\chi^2$ curve is given as a
function of the Higgs boson mass. The blue band indicates the theory
uncertainty from unknown higher-order corrections. The preferred value
for the Higgs-boson mass within the Standard Model, corresponding to the
minimum of the curve, is around 81 GeV, while the 95\% C.L.\ upper limit
(one-sided, corresponding to $\Delta\chi^2 = 2.7$)
obtained from the fit is about 193~GeV. It should be noted that the
result for the Higgs-boson mass from the fit is strongly correlated to
the value of $m_t$. Changing $m_t$ by 5~GeV, corresponding to the
present $1 \sigma$ error, gives rise to a shift in the upper bound for
$M_H$ of about 35\%.

In the above fit results it has been assumed that the Standard Model
provides the correct description of the data. A measure of how well the
Standard Model describes the data is given by the fit probability. This
probability is presently only 1.3\%. It should be noted that the
preference for a light Higgs boson within the Standard Model is not
induced by those observables which significantly deviate from the
Standard Model prediction. Omitting the NuTeV measurement from the fit
leads to almost unchanged results for the fitted parameters, while the
fit probability is improved to 11.4\%. Enlarging the errors of different
measurements entering the effective weak mixing angle at the Z resonance
in order to take into account their spread by almost $3 \sigma$ leads to
a further significantly improved fit probability and a more pronounced
tendency towards a light Higgs boson.

QCD is in a very good shape within the framework of the SM.
Jet production has been measured at HERA for $0 < Q^2 < 10^4$ GeV$^2$
and $4 < E_T < 100$ GeV \cite{foudas}. The jet data have then been used
at HERA I to test QCD and make precise measurements of the gluons and
$\alpha_s$.  The detector and theory systematic errors have been well
understood showing that jet measurements can be made to better than
10\% level if the $E_T$ is high enough (high $Q^2$ is needed also). Low
$Q^2$ regime is rich but require more precise theoretical
calculations.  HERA II will start soon, and one can expect a large
increase of the high $E_T$ high $Q^2$ samples which will enable to measure
the proton PDFs at higher x, measure $\alpha_s$ with better precision, 
BFKL and of course
search for new physics.

\begin{figure}[h]
\begin{center}
  \begin{minipage}{2.3in}\hspace*{.3cm}
\psfig{figure=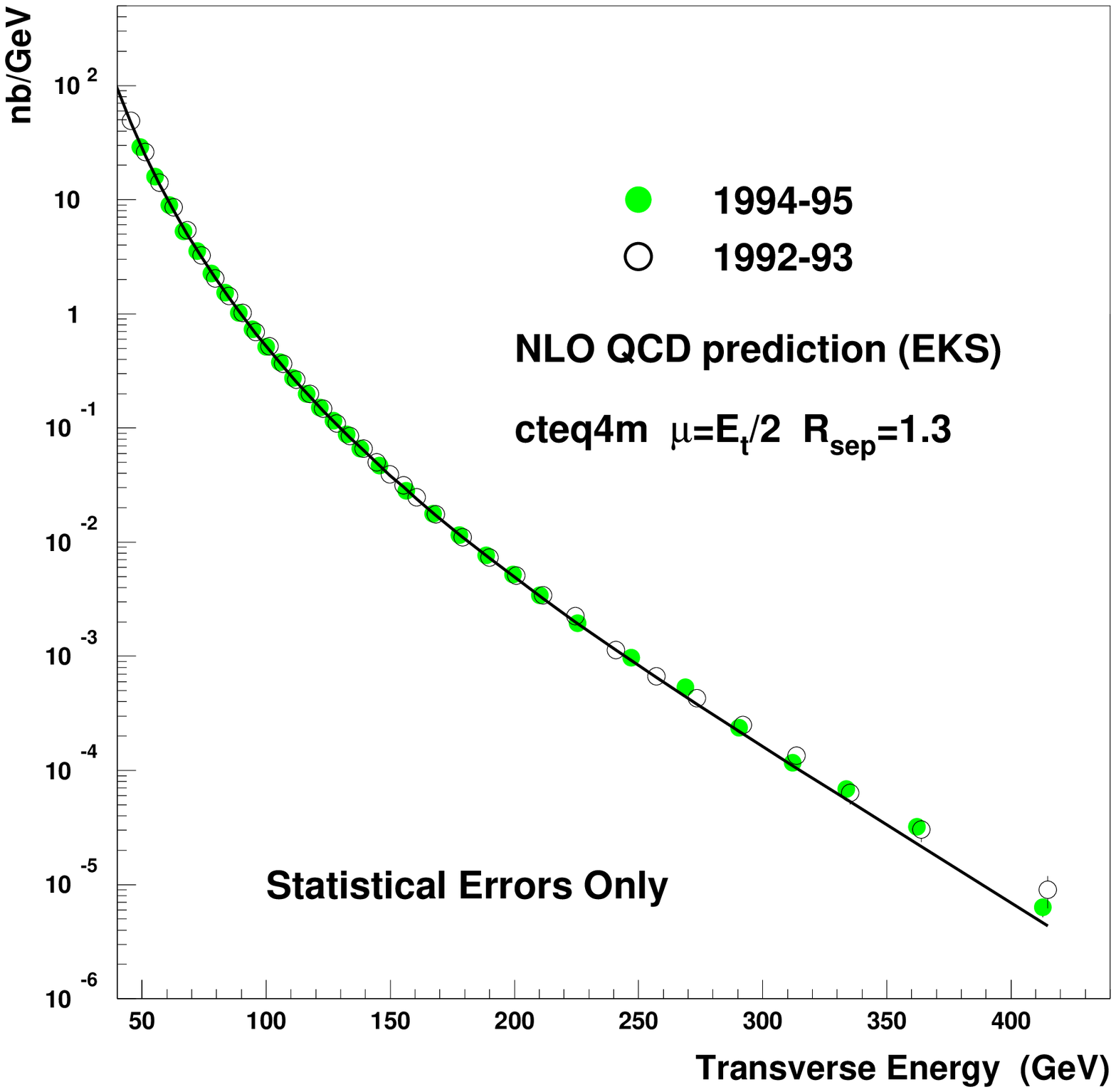,height=2in}\vspace*{-.2cm}
\caption{ {\scriptsize  CDF inclusive jet cross section from Run1B data 
(1994-1995) compared to QCD prediction and to the published Run1A results. }}
    \label{figure5}
  \end{minipage}\vspace*{-.7cm}
\hfill
\hspace*{-.5cm}
  \begin{minipage}{2.3in}\hspace*{.5cm}
\psfig{figure=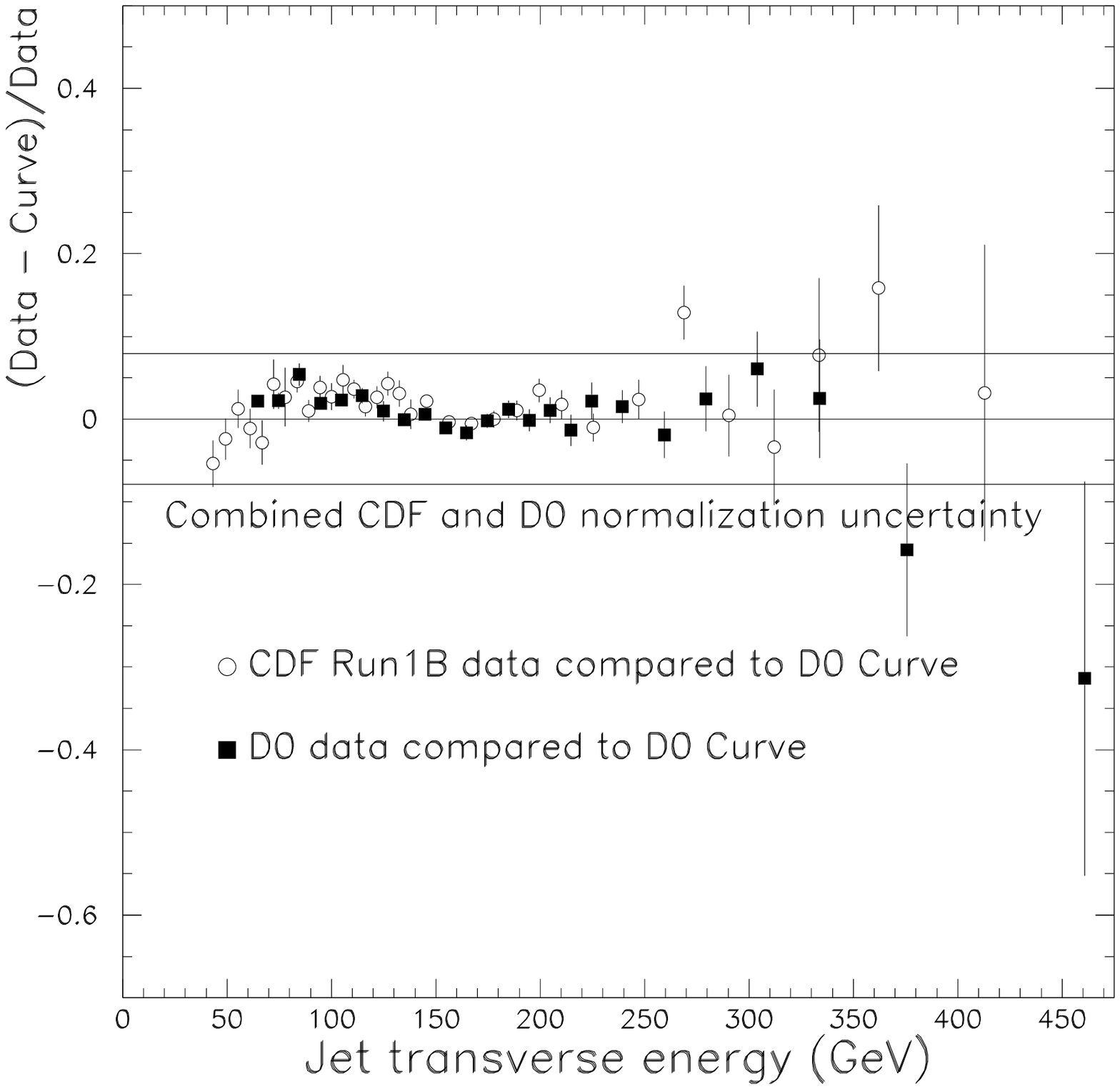,height=2in}
\caption{{\scriptsize Comparison of CDF and D0 data to D0 smooth curve}}
    \label{figure6}\vspace*{.6cm} 
  \end{minipage}
\hfill
\end{center}
\end{figure}

The TeVatron hadron collider provides as well a unique opportunity to
study QCD at the highest energies~\cite{christina}.  The results on jet
production are used extensively to derive new parton distribution
functions and photon data are used to discriminate between different
approaches for understanding their disagreement of the theory with
data relative to photon production at small transverse momentum, 
Figs.~\ref{figure5} and \ref{figure6}.

\section{Top physics}
The top quark was predicted in the Standard Model of electroweak
interactions as a partner of the b-quark in a SU(2) doublet of the
weak isospin, in the third family of quarks~\cite{chris}.  The top
quark was observed at the TeVatron by the CDF and D0 collaboration
during the Run I data taking (1992-1996). The CDF and D0 top mass
averages, obtained from measurements in several channels and based
upon 100 pb$^{-1}$ data from $p\bar p$ collisions at $\sqrt{s}$ =
1.8 TeV which were collected by each experiment in Run I.
The combined
TeVatron measurement of the top quark mass is $M_{top}$ = 174.3 $\pm
3.2$ (stat)$\pm 4.0$ (sys) GeV/c$^2$.  The CDF measurement of the
$t\bar t$ cross section (assuming $M_{top}$ = 175 GeV/c$^2$) is
$\sigma_{t\bar t}$ = 6.5$\pm^{1.6}_{1.4}$ pb and the D0 value 
(assuming $M_{top}$ = 172.1 GeV/c$^2$) is $\sigma_{t\bar t}=5.9\pm 1.7$~pb.

All mass measurement techniques assume that each selected event
contains a pair of massive objects of the same mass (top and anti-top
quarks) which subsequently decay as predicted in the SM.  A variety of
fitting techniques use information about the event kinematics.  A
one-to-one mapping between the observed leptons and jets and the
fitted partons is assumed~\cite{chris}.  Of course, it is
assumed that the selected sample of events contains just the $t \bar{t}$
events and the SM background. 
This is the simplest and the most natural hypothesis since top quark is expected in the SM.
On the other hand, the samples of $t \bar t$ and single top candidates are
among the best places to look for new physics. Because of the top
quark mass being large, event selection cuts in top analyses are
practically identical to those applied in many analyses looking for
physics beyond the SM (Supersymmetry, Technicolor, Leptoquarks, etc). 
Both CDF and D0 made numerous comparisons between
various distributions of the reconstructed top quarks, and
especially those of the $t \bar t$-system, with the SM
predictions. No significant disagreements were found.

\par 

The increased integrated luminosity expected at the TeVatron Run II 
(about a factor of 20 
in respect to Run I), combined with improvements to CDF and D0 detectors and
larger $t \bar t$ cross section, will allow the experiments to collect a 
number of reconstructed top events  20--70 times larger than in Run I, 
depending on the
final state and tagging requirements.  The systematic effects will
dominate uncertainties in the measurements of $\sigma_{tt}$ and M$_t$.
Both experiments estimate that the error on M$_t$ will reach $\Delta
M_{top}$= 2--3 GeV/c$^2$ (compared with 7 GeV/c$^2$ in Run I).  The $t
\bar t$ cross section should be measured with an error of about 8\%
(about 30\% in Run I).  Analysis of single top production offers a
direct access to the $Wtb$ vertex and should allow the measurement of
the $|V_{tb}|$ element of the Cabibbo-Kobayashi-Maskawa matrix.  Anomalous
couplings would lead to anomalous angular distributions and larger
production rates.  The expected SM cross sections are of the order of
1--2 pb. Of course the increased statistics will allow the TeVatron 
experiments to 
finally test the underlying hypothesis that the top
candidate events are just the $t \bar t$ events and not events from new
physics~\cite{Mumbai}.

\section{The quest for the Higgs boson}

The Higgs mechanism is one of the basic ingredients of the Standard
Model theory of fundamental interactions, allowing the introduction of
masses for the observed particles, without violating the local gauge
invariance. Still the Higgs boson is probably the most elusive
particle to be found.  Within the SM a single neutral scalar is
predicted, whose mass is an arbitrary parameter, although unitarity
of the model imposes an upper limit of about 1 TeV.
Precision electroweak measurement indicate that the mass of the
Standard Model Higgs boson should be around 81 GeV/c$^2$
and the 95\% C.L. upper limit is set at 193 GeV/c$^2$. 
The Higgs boson has still to be found.  It has been searched 
for extensively at
LEP. In particular, in the years 1998-2000 the four LEP collaborations
have collected 2465 pb of data from $e^+e^-$ collisions at $\sqrt{s}$
between 189 and 209 GeV.  The SM Higgs boson is
produced via Higgs-strahlung or vector boson fusion and decays mainly
into $b \bar b$. The searches at LEP are based on the following
topologies: fully hadronic decay ($H\to b \bar{b} Z\to q\bar{q}$),
decay of the Z into $\nu \bar \nu$ and decays of the Z into $l\bar l$
or $\tau \bar{\tau}$. The fully hadronic channel has the largest cross
section and the possibility of detection is dominated by the b-tagging
capabilities of the detectors.

The Higgs search at the highest c.m.\ energies at LEP resulted in an
excess of signal-like events above the background expectations with a
statistical significance of about $1.7 \sigma$, compatible with the
production of a Standard Model Higgs of about $M_H = 116$~GeV. The
exclusion bound on the Standard Model Higgs obtained by combining all
LEP data is $M_H > 114.4$~GeV at 95\% C.L.~\cite{higgs2}.

The quest for Higgs will now move to the TeVatron, where a new data taking
phase started in March 2001 (Run II)~\cite{petteni}.
At the TeVatron Run II the $gg\to H$ production mode
dominates over all mass ranges, but the huge irreducible 
QCD background makes it impossible to use this production channel
for a measurement.
So, for low Higgs mass ($M_H$ $<$ 130 GeV/c$^2$) the 
associated production with a vector boson
($W$ or $Z$) and the subsequent 
decay into $b \bar b$ 
is the most promising channel, with an estimated cross section
production of order 0.1~pb.  The double b-tagging of the 2 jets coming
from the Higgs decay, together with the signature of the additional
boson helps to discriminate from the background.  From the trigger
point of view, channels with one high $p_T$ lepton coming from the
vector boson decay are not a concern, since the rate can be easily
controlled. On the other hand channel where the vector boson decays
into quarks (W/Z) or neutrinos (Z) have an higher branching ratios
and trigger strategies need to be devised to control data taking
rates. It has been shown in preliminary studies~\cite{snowmass} that a
trigger strategy based on the use of SVT tracks is crucial in
selecting a sample enriched in heavy flavors, keeping the rate at a
reasonable level.  Improved offline b-tagging efficiency (a factor
1.3 is already achieved only due to the increased geometrical coverage
of the silicon detector) will then help in discriminating signal from
background.

\begin{figure}[h]
\begin{center}
\psfig{figure=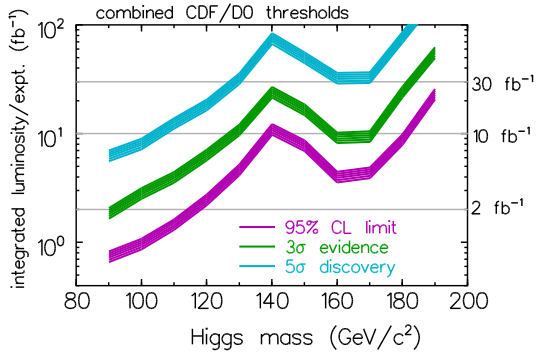,height=2.0in}\vspace*{-.4cm}
\caption{ {\scriptsize
     Projected discovery/exclusion regions for SM Higgs as function of 
luminosity at Run II}}
    \label{run2}
\end{center}\vspace*{-.5cm}
\end{figure}

In Fig.~\ref{run2} the expected discovery reach in Run II for the 
Standard Model 
Higgs boson from the study carried out during the Run II workshop at 
Fermilab \cite{fnal_runII}.
Based on a simple detector simulation,  
the integrated luminosity necessary to discover the SM Higgs in the 
mass range 100-190 GeV was estimated.
The first phase of the Run II Higgs search, 
with a total integrated luminosity of 2 fb$^{-1}$ per detector, 
will provide a 95\% C.L. exclusion sensitivity comparable to 
that one obtained at LEP.
With 10 fb$^{-1}$ per detector, this exclusion will extend up to 
Higgs masses of 180 GeV, and a tantalizing 3 sigma effect will be visible 
if the Higgs mass lies below 125 GeV. 
With 25 fb$^{-1}$ of integrated luminosity per detector, evidence for 
SM Higgs production at the 3 sigma level is possible for Higgs masses up 
to 180 GeV. 
However, the discovery reach is much less impressive for achieving a 
5 sigma Higgs boson signal. Even with 30 fb$^{-1}$ per detector, 
only Higgs bosons with masses up to about 130 GeV can be detected with 
5 sigma significance. 

\section{Beyond the Standard Model}
\label{sec:beyond}
Although the Standard Model is remarkably successful in describing
all experimental results, there are theoretical arguments to believe
that it is only a low-energy effective theory.
Searches for ``new physics'', which could reveal the true fundamental 
theory of particles and interactions are among most interesting
subjects in present and future colliders.
Large variety of results was presented at this conference.
\subsection{SUSY}
Supersymmetry (SUSY) is  believed to be 
the best motivated candidate for a theory 
beyond the Standard Model.
It solves the hierarchy problem and provides a framework for 
consistent gauge unification.  
Supersymmetry predicts that for each fermion and gauge boson of the SM
a supersymmetric partner with spin different by 1/2 unit exists.
This opens a wide field for discoveries at present and future colliders.

\begin{figure}[h]
\begin{center}
\epsfig{figure=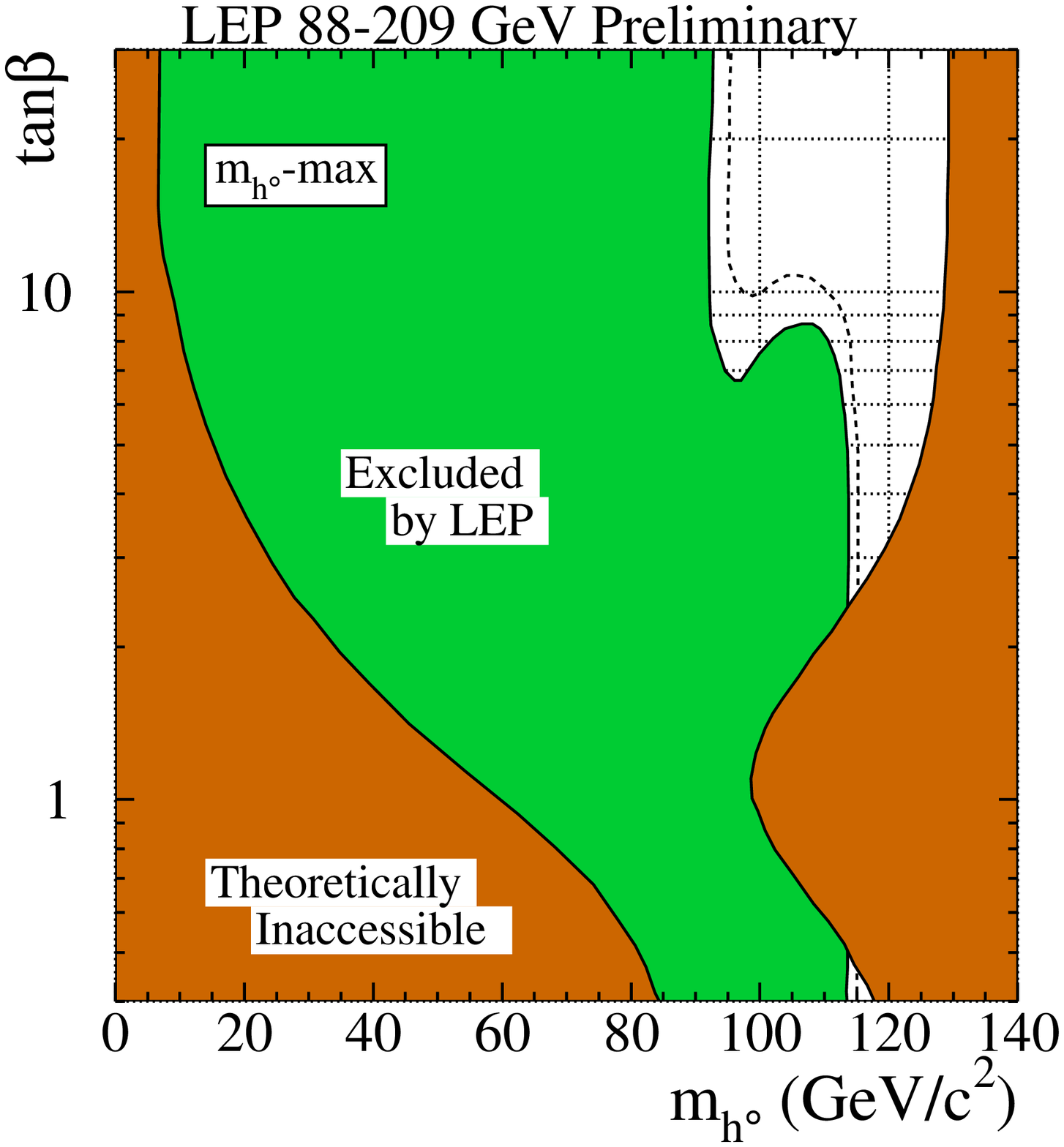,height=6cm,clip=}
\epsfig{figure=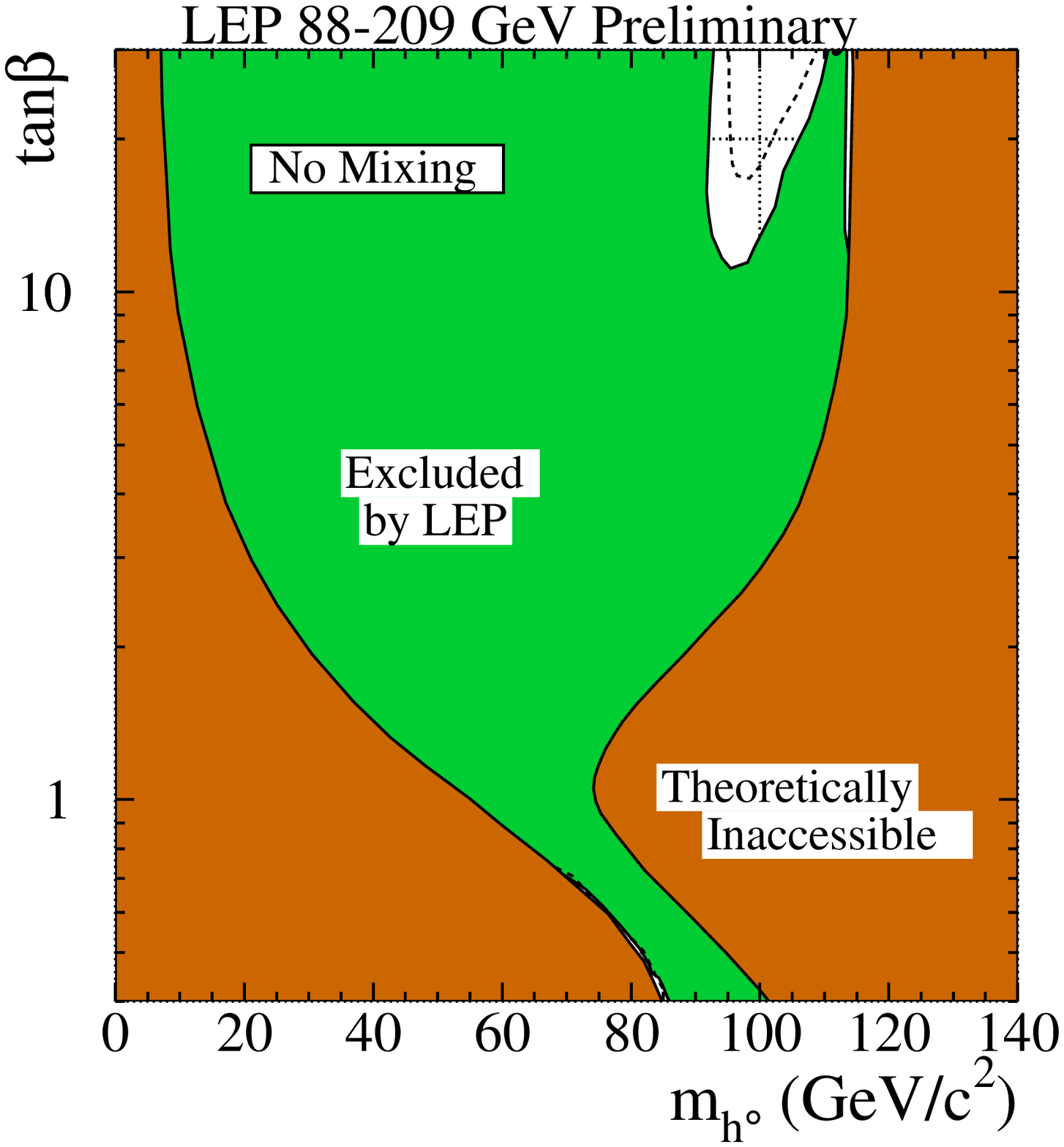,height=6cm,clip=}
\end{center}
\vspace*{-.5cm}
 \caption{LEP exclusion limits in $(m_h, \tan \beta)$ plane
 for the $m_{h}^{\mbox{max}}$ (left plot) and the no-stop-mixing (right plot)
MSSM benchmark scenario \cite{carena}.}
  \label{fig:susy_h}
\end{figure}

The Minimal Supersymmetric extension of the Standard Model (MSSM)
requires two Higgs doublets, giving rise to five physical Higgs bosons,
$h$, $H$, $A$, $H^{\pm}$. In contrast to the Standard Model, the mass of
the lightest CP-even Higgs boson in the MSSM is not a free parameter,
but can be predicted within the model. This leads to the tree-level
upper bound of $m_h < M_Z$, which however is affected by large radiative
corrections. Taking into account corrections up to the two-loop level,
an upper bound of about $m_h < 135$~GeV can be established 
\cite{Weiglein}. This bound
is valid for $m_t = 175$~GeV and is shifted upwards by about 5~GeV if 
$m_t$ is shifted by $+5$~GeV. The exclusion limits obtained from the
Higgs search at LEP are shown in the plane of $m_h$ and $\tan \beta$,
the ratio of the vacuum expectation values of the two Higgs doublets,
for two MSSM benchmark scenarios \cite{Weiglein} 
in Fig.~\ref{fig:susy_h} \cite{lep_susy_h}. 
Note that in these benchmark scenarios $m_t = 174.3$~GeV is used. A
shift in $m_t$ would significantly affect the upper bound on $m_h$ as
function of $\tan\beta$ (the ``theoretically inaccessible'' region for high
$m_h$ values in the plots). The Higgs searches at the TeVatron can probe a
significant part of the MSSM parameter space. In fact, the discovery of
a Higgs boson with non-SM couplings could be a first sign of
supersymmetry. While the upper bound on $m_h$ can be somewhat relaxed in
non-minimal SUSY models (up to about 200~GeV), the prediction of a light
Higgs boson is generic to all SUSY models which stay in the perturbative
regime up to the GUT scale.

LEP covered most of the MSSM parameter space, but the Higgs boson
was not found. 

\begin{figure}[h]
\begin{center}
\epsfig{figure=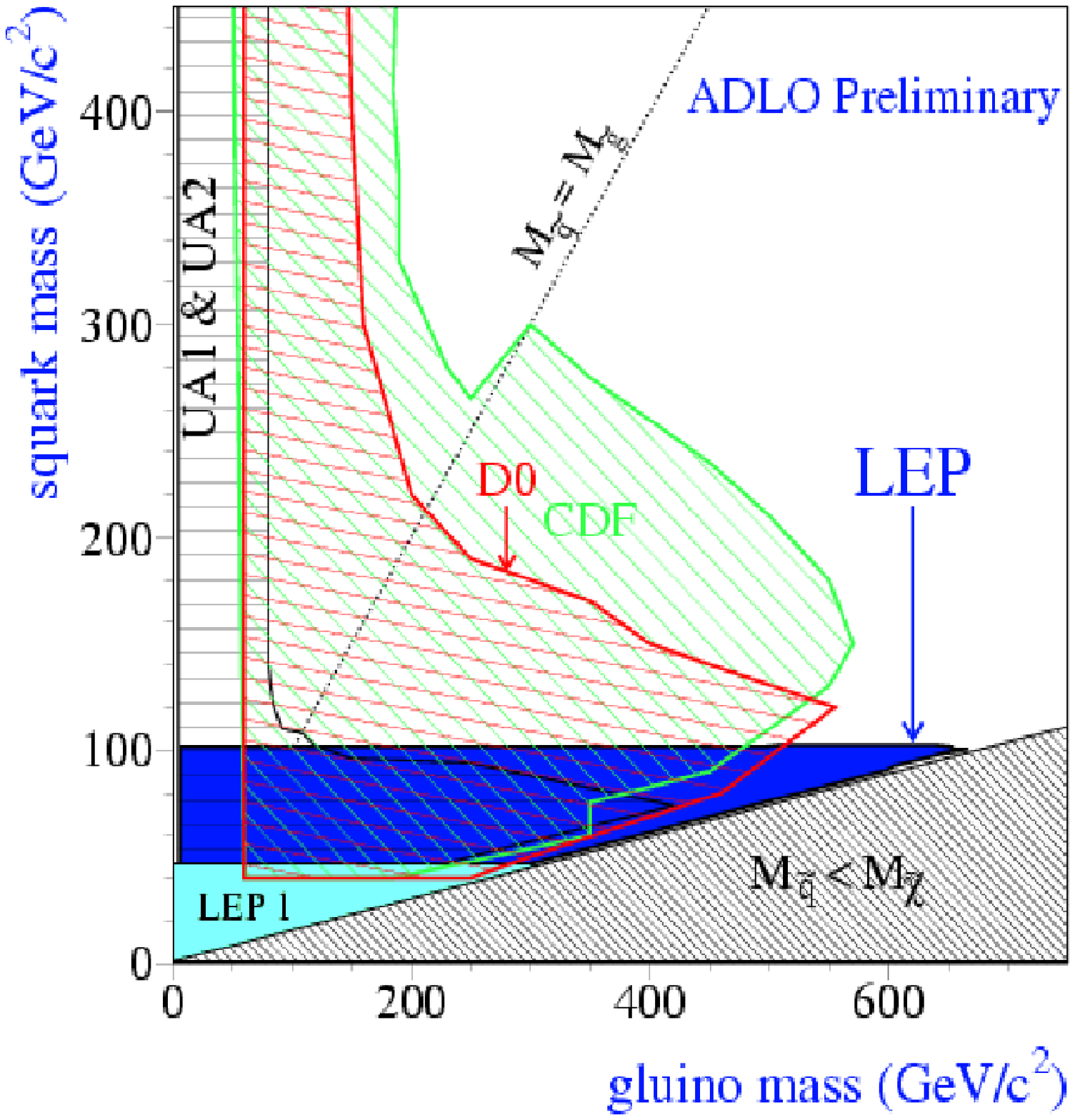,height=6cm,clip=} 
\epsfig{figure=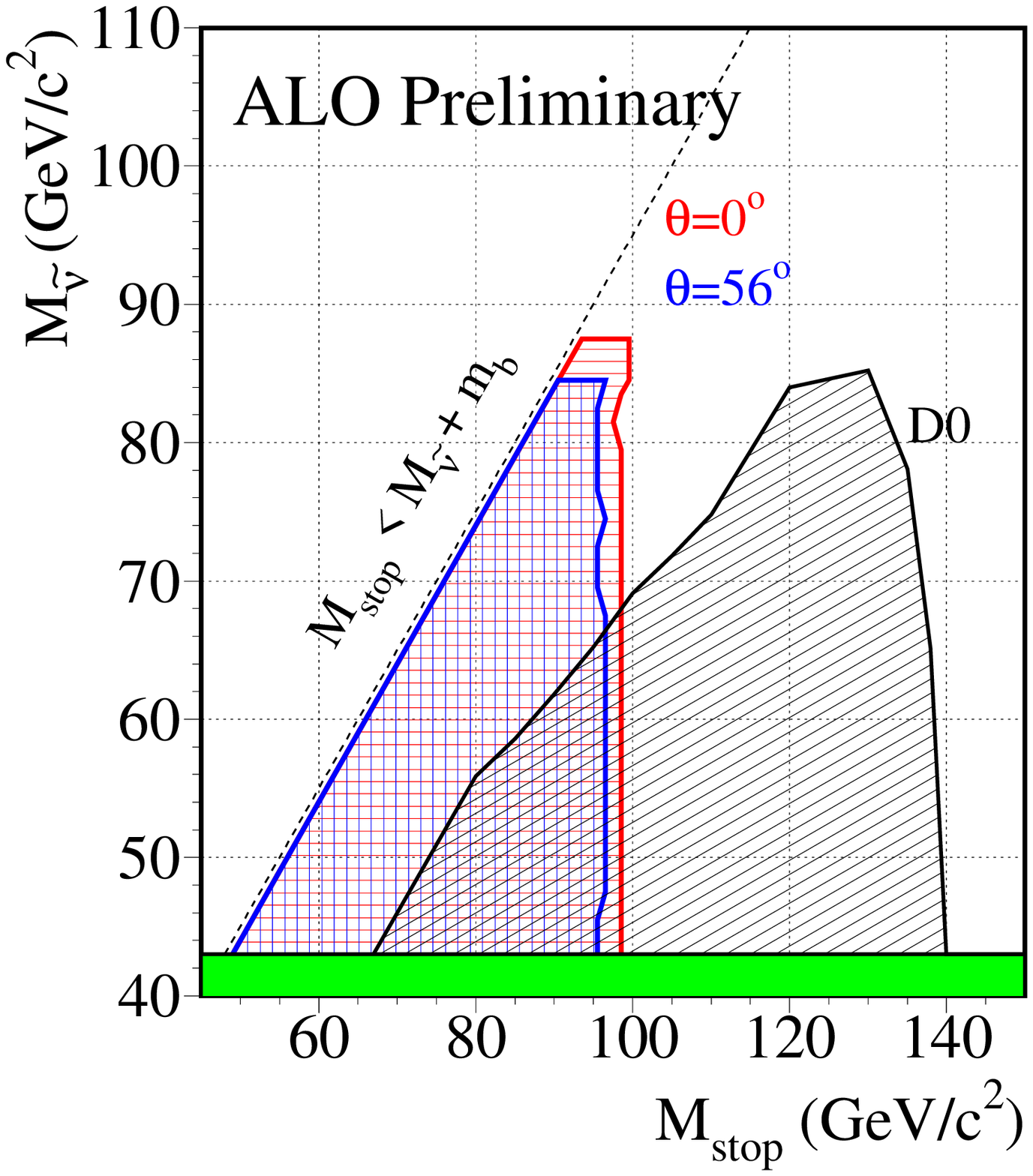,height=6cm,clip=}
\end{center}\vspace*{-.6cm}
 \caption{Comparison of gluino and squark mass limits (left plot) and
stop and snutrino mass limits (right plot) from TeVatron and LEP.}
  \label{fig:susy_q}
\end{figure}

LEP searches for other supersymmetric particles have shown no
evidence for a signal. Therefore exclusion bounds under
certain model assumptions can be derived and e.g.
stop, selectron and chargino masses are excluded up to
96, 99.4 and 103.5 GeV, respectively~\cite{lep_susy_part}.
The Lower 95\% C.L. limit on the neutralino LSP mass is about 45 GeV.
The LEP squark mass limits are compared with new CDF squark and gluino
search results in Fig.~\ref{fig:susy_q} (left).
New analysis of CDF data  exclude gluino masses  below about 180 GeV 
\cite{cdf_susy_new}.
Complementary results are obtained by LEP and TeVatron
as for the search for stop decaying to sneutrino
($\tilde{t} \rightarrow b l \tilde{\nu}$), as shown in
Fig.~\ref{fig:susy_q} (right).
Stop masses up to about 140 GeV are excluded by D0 under this 
assumption \cite{d0_stop}, whereas
LEP experiments give limits for models with low stop-sneutrino mass difference.

\begin{figure}[h]
\begin{center}
\epsfig{figure=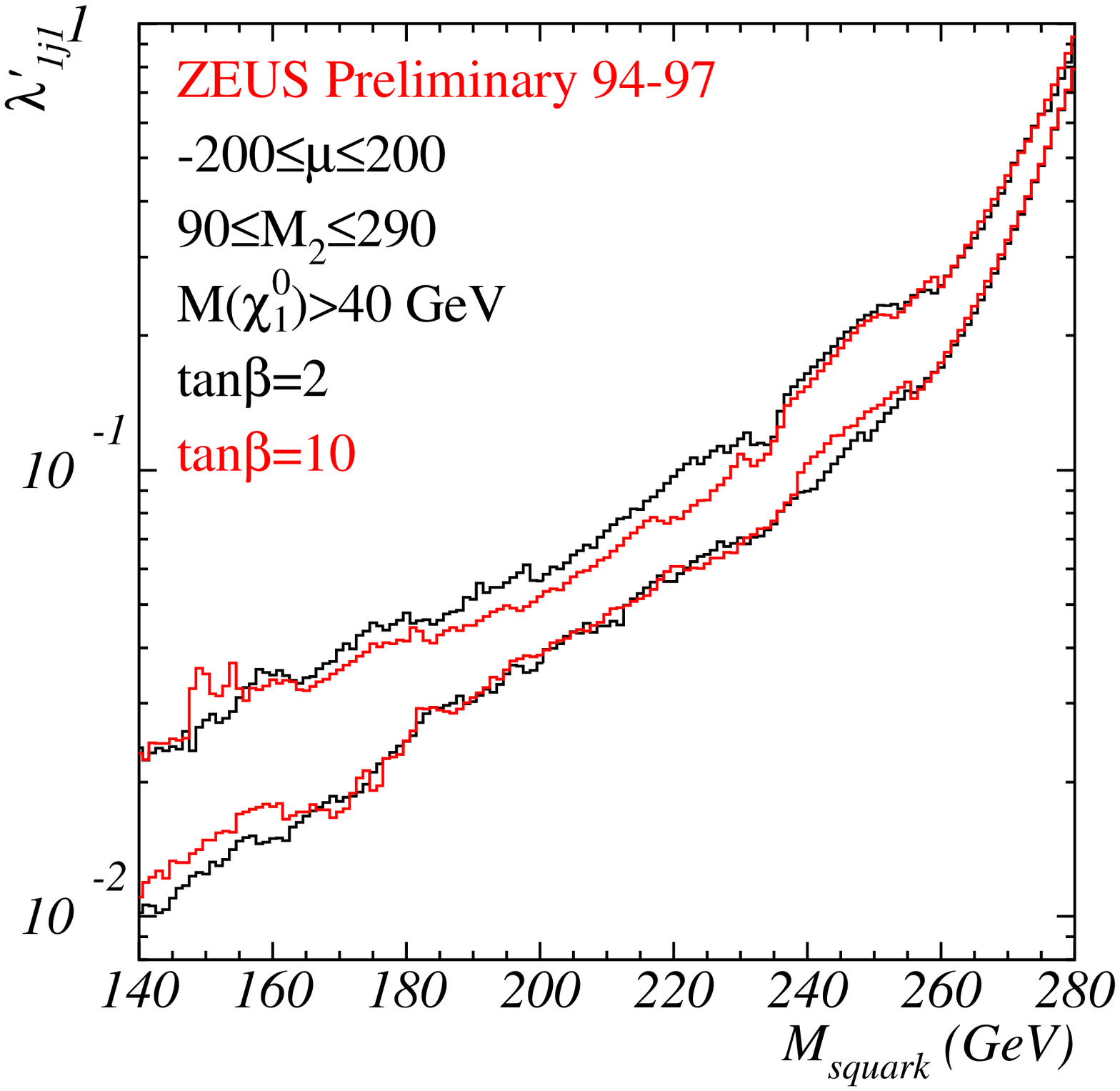,height=6cm,clip=}  
\epsfig{figure=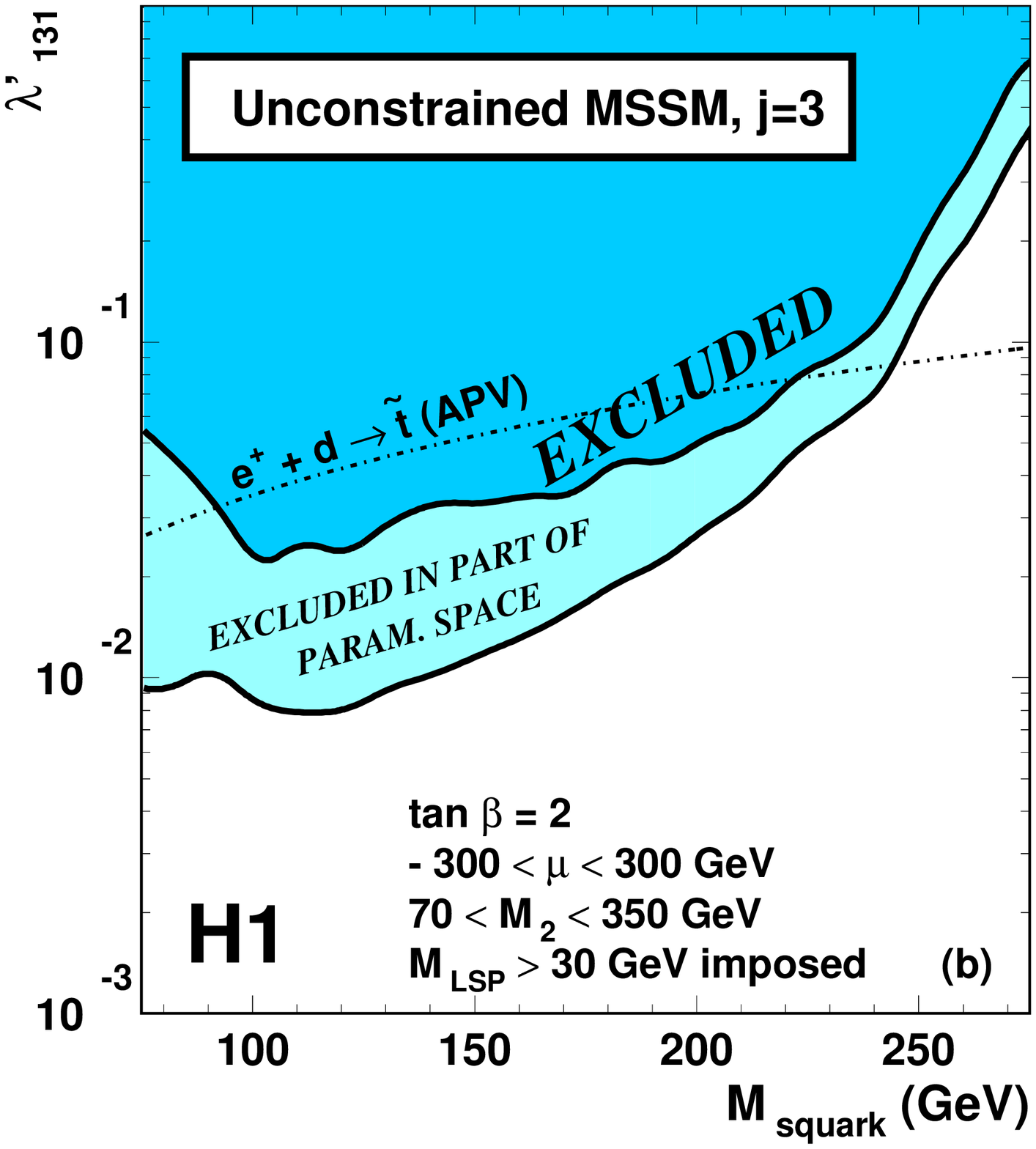,height=6.4cm,clip=}
\end{center}\vspace*{-.6cm}
 \caption{ZEUS and H1 limits on the coupling $\lambda'_{1j1}$ 
as a function of the squark mass in the $R_P$-violating SUSY.
Coupling values above the upper curve are excluded on 95\% C.L. for all
values of $\mu$ and $M_2$, whereas those above lower line are
excluded only in a part of parameter space. }
  \label{fig:susy_rp}
\end{figure}

Both HERA collaborations searched for squark production~\cite{hera_mssm}
and investigated different possible decay channels.
In the SUSY models with $R_P$-violation resonant 
squark production is possible at HERA~\cite{hera_rp}.
Resulting limits on the $R_P$-violating coupling $\lambda'_{1j1}$ 
as a function of the squark mass are shown in Fig.~\ref{fig:susy_rp}.
For coupling of electro-magnetic strength ($\lambda'_{1j1}$=0.3)
squark masses up to 260 GeV are ruled out.

\subsection{Large Extra Dimensions}
Another possibility of solving the hierarchy problem 
of the Standard Model has been proposed recently.
The problem is avoided
if additional compactified spacial dimensions
are introduced.
If the extra dimensions are large 
the effective Planck scale $M_S$ can be in the TeV 
range~\cite{add,Randall:1999ee}.

The propagation of graviton in extra dimensions $n$
can lead to effects observable at high energy colliders.
The signature for real graviton production is 
single vector boson or monojet with large transverse momentum
and large missing transverse momentum due to the escaping graviton.
For $n=2$ limits on  $M_S$ between 1 and 1.4 TeV are set by 
LEP experiments from search for $e^+ e^- \rightarrow \gamma G$ 
events \cite{lep_ed_dir}.
For $n \ge 5$ best limits on $M_S$ of the order of 600--650 GeV 
(slowly decreasing with $n$) are obtained from monojet events
at the TeVatron \cite{tev_ed_dir}.

\begin{figure}[h]
\begin{center}
\epsfig{figure=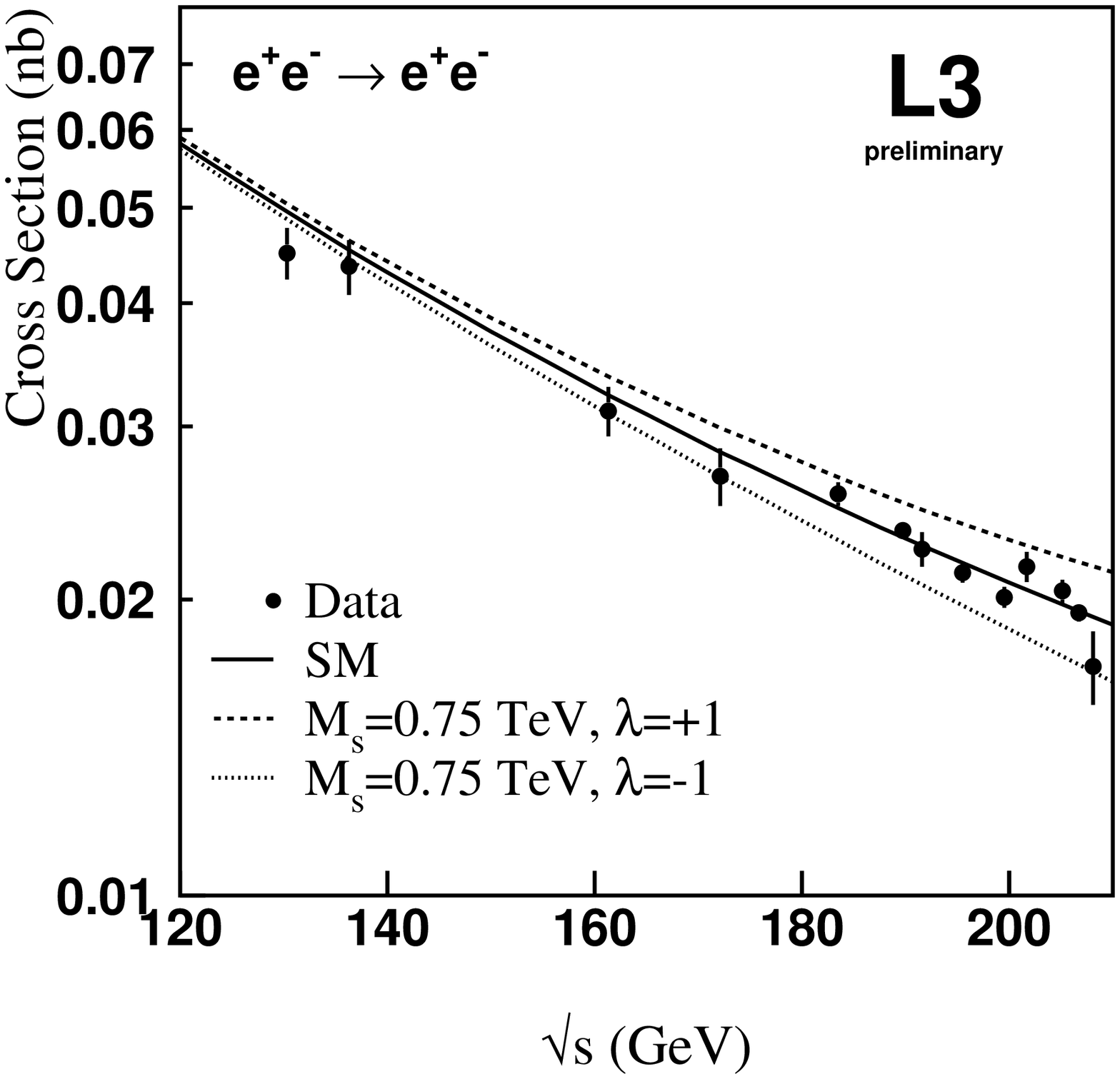,height=6.3cm,clip=}  
\epsfig{figure=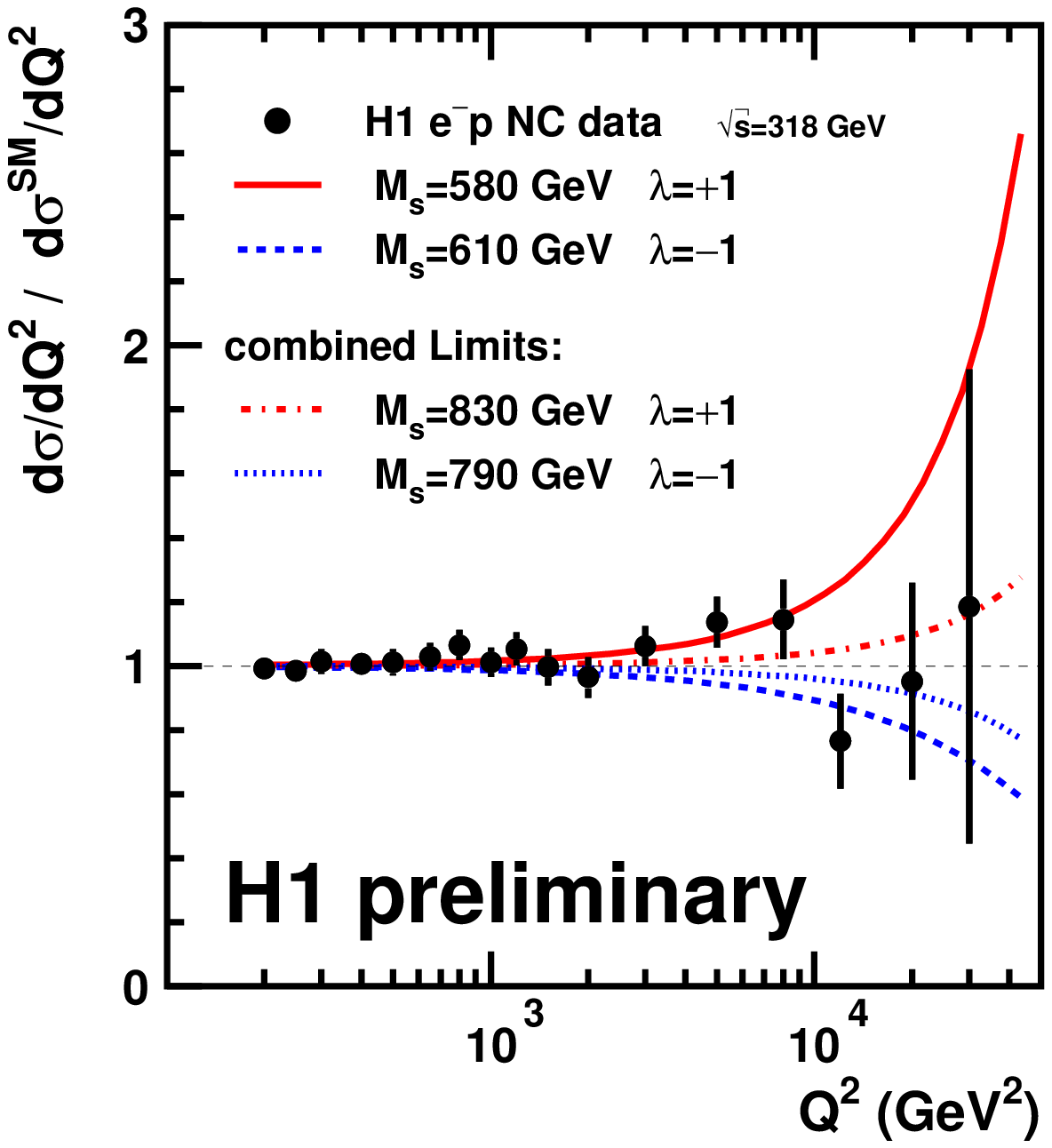,height=5.7cm,clip=}
\end{center}\vspace*{-.5cm}
  \caption{Left: comparison of the Bhabha cross section measured at L3
           for $\sqrt{s} = $130 -- 209 GeV 
           with SM predictions and predictions of the ADD model.
           Right: H1 $e^-p$ data 
           compared with 95\% exclusion limits for 
           the effective Planck mass scale in models with large extra 
           dimensions.
           }
  \label{fig:led}\vspace*{.3cm}
\end{figure}

Virtual graviton exchange could also contribute to fermion-pair and
boson-pair production at LEP and TeVatron, as well as to NC DIS
at HERA.
Fig.~\ref{fig:led} (left) shows the cross section for Bhabha scattering
$e^+ e^- \rightarrow e^+ e^- $ measured by L3, compared with the SM
predictions and predictions of the ADD (Arkani--Hamed, Dimopoulos, Dvali) 
model~\cite{add_lep_pred}.
From analysis of $e^+ e^- $ and $\gamma \gamma$ events
combined LEP limit on $M_S$ is 1.13 (1.39) TeV, for positive
(negative) coupling \cite{lep_ed_vir}.
Similar analysis of $e^+ e^- $ and $\gamma \gamma$ events
at the TeVatron resulted in the 95\% C.L. limit on $M_S$ 
of 1.1 (1.0) TeV \cite{tev_ed_vir}.
A combined analysis of the $e^+ p$ and $e^- p$ data at HERA resulted
in $M_S$ limits of about 0.8 TeV from H1 and ZEUS \cite{hera_ci}.
H1 $e^-p$ data 
compared with 95\% exclusion limits for  $M_S$ is shown in 
Fig.~\ref{fig:led} (right).

\subsection{Leptoquarks}
The striking symmetry between quarks and leptons suggests that
there could exist a more fundamental relation between them.
Such lepton-quark  "unification"  is achieved for example 
in different theories of grand unification 
and in compositeness models.
Quark-lepton bound states, called leptoquarks, carry both colour and 
fractional electric charge and a lepton number.
A general classification of leptoquark states used in many analyses
has been proposed by \cite{brw}, where
7 scalar and 7 vector leptoquarks are considered.

\begin{figure}[h]
\begin{center}
\epsfig{figure=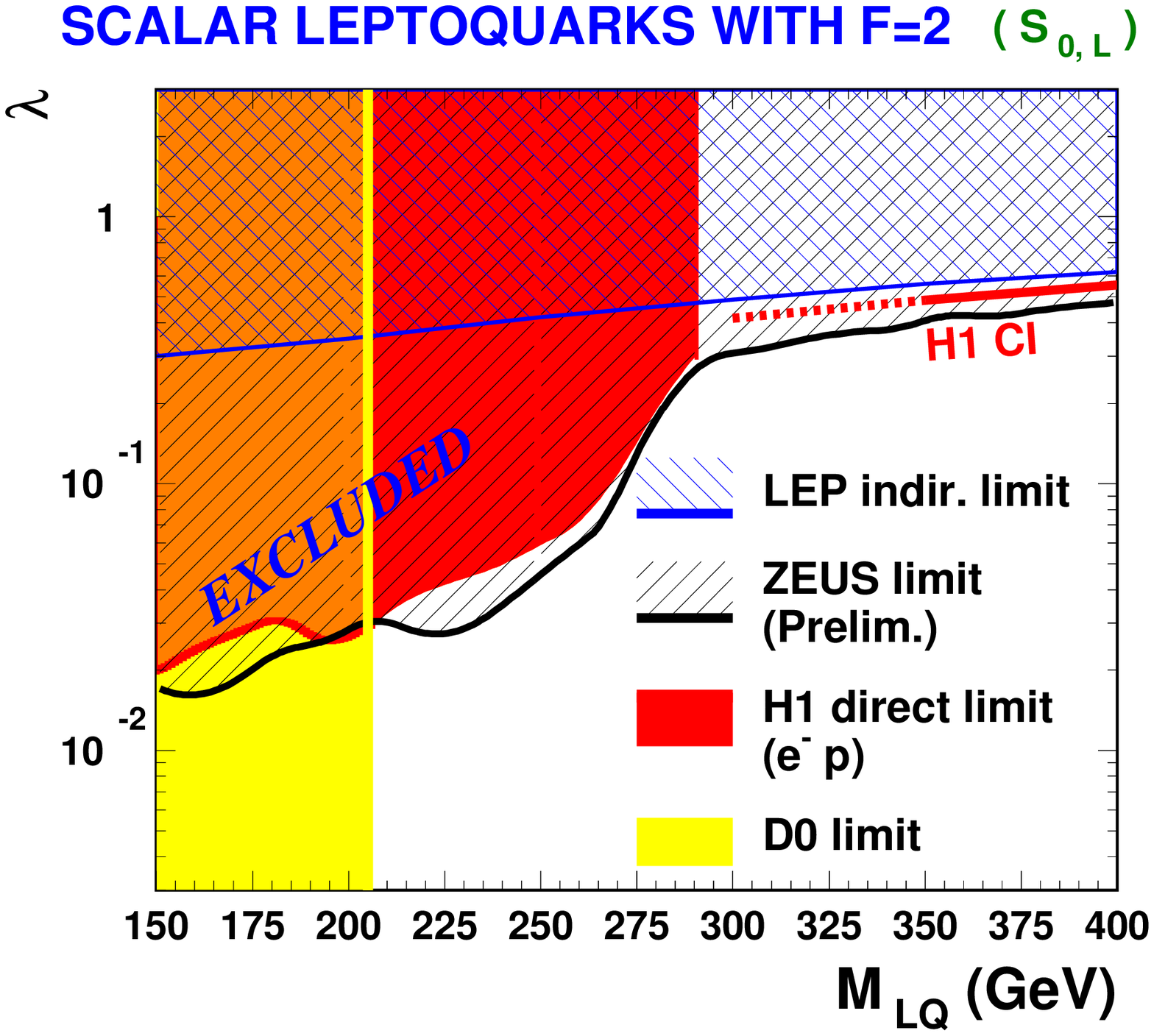,width=6cm,clip=}  
\epsfig{figure=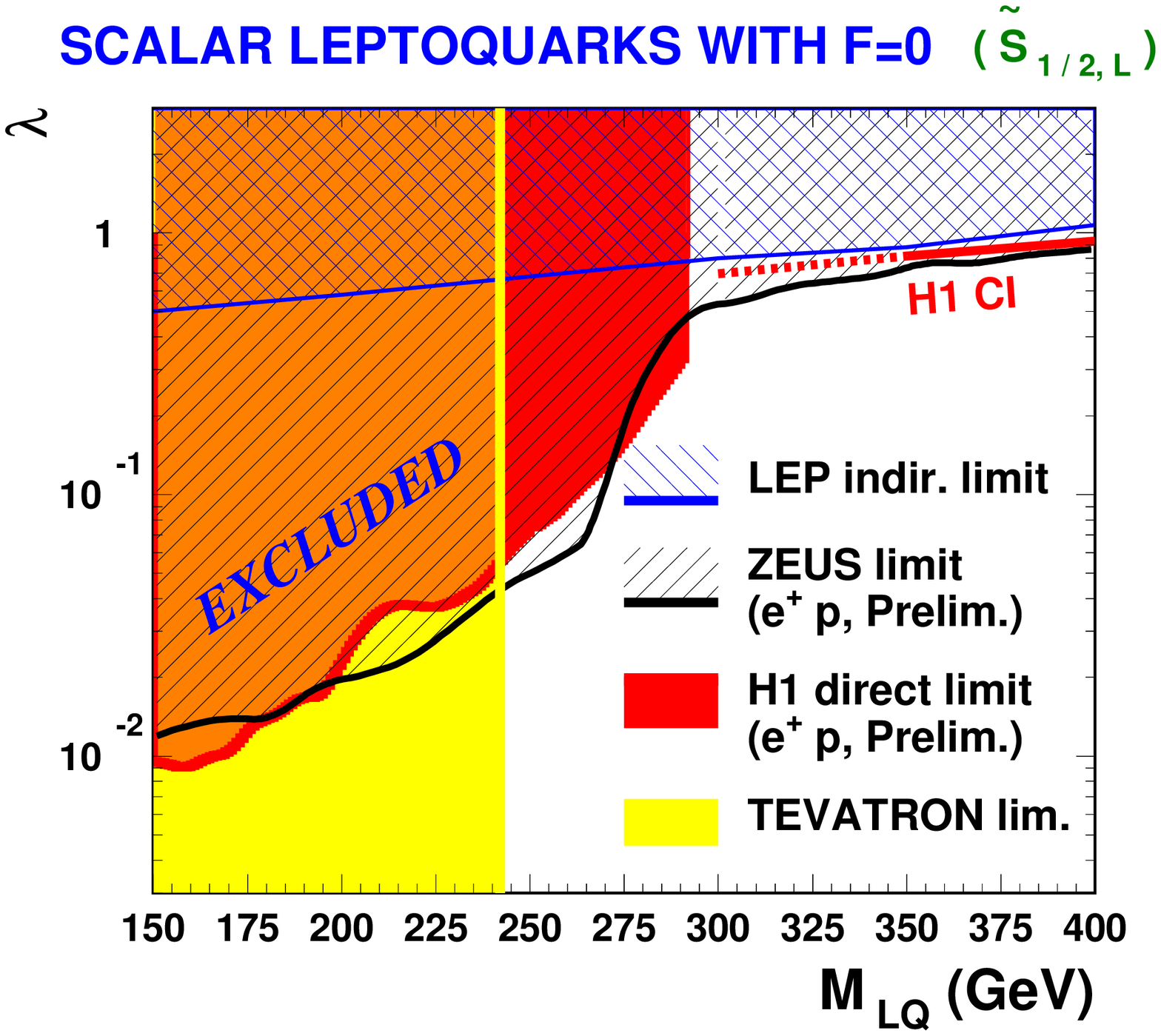,width=6cm,clip=}
\end{center}\vspace*{-.5cm}
 \caption{Comparison of scalar leptoquark limits from HERA, TeVatron and LEP.}
  \label{fig:sztuk}
\end{figure}

At HERA leptoquarks could be resonantly produced via fusion of
the incoming lepton (electron or positron) and a quark (or antiquark) 
from the proton, with subsequent decay into $e^\pm$-quark or 
$\nu (\bar{\nu})$-quark.
For large leptoquark masses, $m_{LQ} > \sqrt{s}$ the $t$-channel leptoquark 
exchange and the interference with SM processes become important.
Both experiments H1 and ZEUS see no evidence for a resonant LQ production
or cross section deviations due to high mass LQ exchange \cite{hera_lq}.
Fig.~\ref{fig:sztuk} compares the H1 and ZEUS results for 
two scalar leptoquarks with results from LEP and TeVatron.
The combined analysis of CDF and D0 data exclude scalar leptoquark
masses below 242 GeV, for leptoquarks decaying to electron-quark only
($\beta = 1$) \cite{tev_lq}.
This limit is based on a 
leptoquark pair-production in strong interactions
and is independent on the leptoquark Yukawa coupling $\lambda$.
Included in Fig.~\ref{fig:sztuk} are also indirect limits from LEP, 
based on the measurement of $q \bar{q}$ production cross section, 
which is sensitive to the virtual $t$-channel leptoquark 
exchange \cite{lep_ff_pair}.
In the high mass region, beyond the kinematical limits, 
HERA and LEP provide comparable limits.

\subsection{Other searches}
Effects  coming from "new physics" at  high energy scales
can be described in the most general way by  
four-fermion contact interactions.
This includes the possible existence of second generation heavy weak bosons, 
heavy leptoquarks as well as electron and quark compositeness.
At HERA, $eeqq$ contact interactions would modify NC DIS cross sections
at high $Q^2$. 
Since no deviations from SM predictions are found, exclusion limits
on the compositeness scale $\Lambda$ of 1.8 to 7.0 TeV are set by
H1 and ZEUS experiments, based on the data collected 
in 1994-2000 \cite{hera_ci}.
Comparable limits are also obtained from measurement of Drell-Yan 
lepton-pair production at the TeVatron~\cite{tev_ci} and hadronic
cross section and charge asymmetries at LEP \cite{lep_ff_pair}.
Limits on four-lepton contact interactions $eell$ set by
LEP experiments range from 8.5 to 26.2 TeV. 

\begin{figure}[h]
\begin{center}\hspace*{-.5cm}
\epsfig{figure=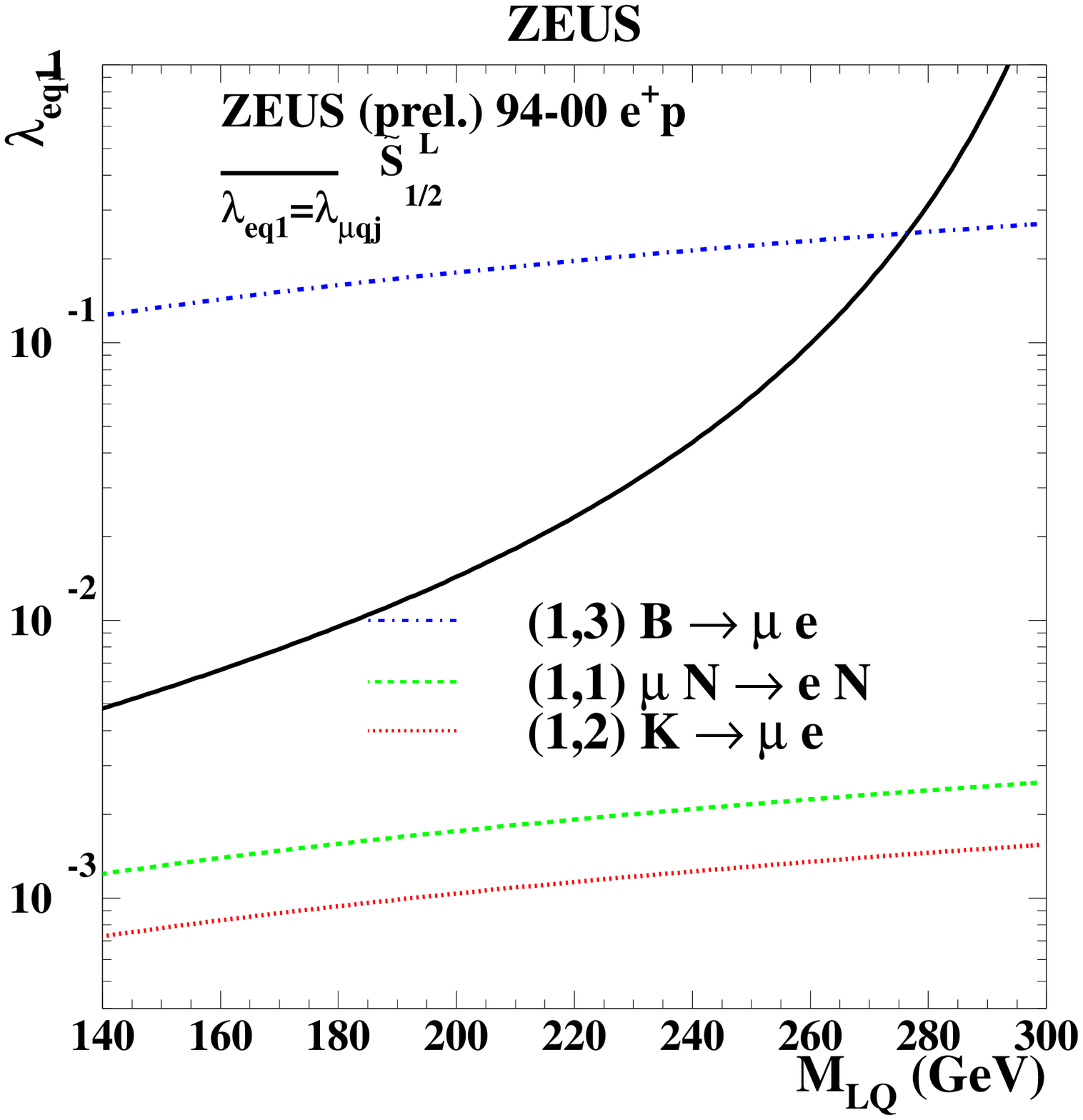,width=6cm,clip=}  
\epsfig{figure=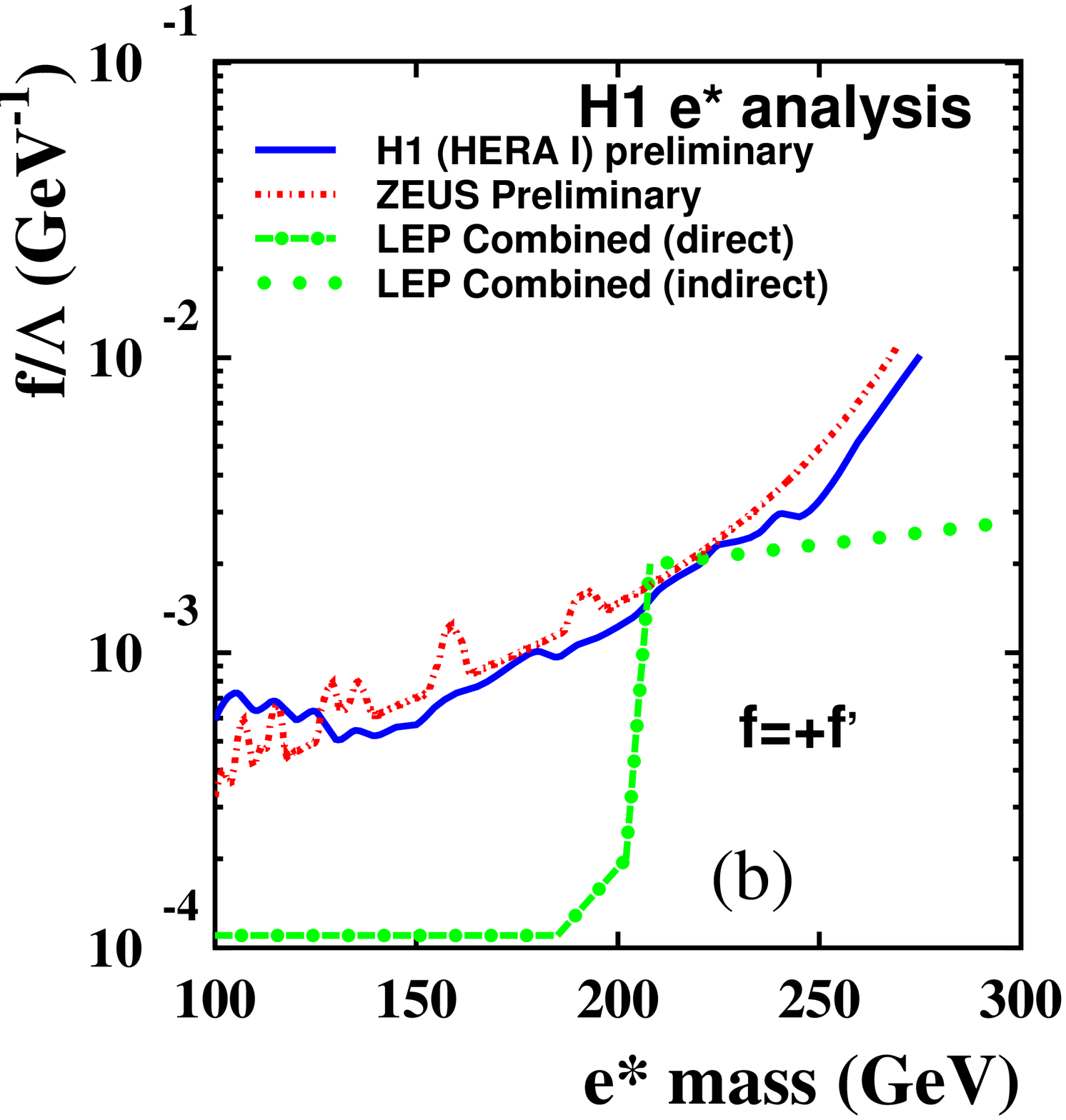,width=5.4cm,clip=}
\end{center}\vspace*{-.5cm}
  \caption{Left: ZEUS 95\% C.L. upper limit on the LFV leptoquark
Yukawa coupling, for  $S^L_{1/2}$ leptoquark decaying into $\mu q$.
Right: H1, ZEUS and LEP upper limits on $f/\Lambda$
           for $e^\star$.}
  \label{fig:lfv+estar}
\end{figure}

Observation of the Lepton-Flavour Violation (LFV) in the neutrino
oscillations suggests that LFV processes could also be observed in the
charged lepton sector.
Both H1 and ZEUS experiments searched for events with high-transverse-momentum
muon or tau production, $e p \rightarrow \mu (\tau ) X$.
LFV at HERA  could be due to the $s$-channel production or 
$u$-channel exchange of a leptoquark coupling to different
lepton and quark generations.
No events consistent with LFV LQ production or exchange were found
\cite{hera_lfv}.
Fig.~\ref{fig:lfv+estar} (left) shows the upper limit on the LFV leptoquark
Yukawa coupling, for $S^L_{1/2}$ scalar leptoquark decaying into $\mu q$.
HERA results are competitive with those from low energy experiments
when heavy quarks are involved.
Assuming the Yukawa coupling $\lambda_{eq1} = 0.3$ LFV leptoquarks with masses
up to about 300~GeV can be excluded at HERA.


The observation of heavy excited fermions would be a clear evidence for
fermion substructure.
H1 and ZEUS experiments reported results from excited electron 
and excited neutrino searches at HERA \cite{hera_fstar}.
Decay channels involving $\gamma$, $Z$ and $W$ boson emission were
considered.
No excess of data events over the expected background has been observed.
Limits on the excited electron coupling over 
the compositeness scale ratio, $f/\Lambda$ 
are shown in Fig.\ref{fig:lfv+estar} (right plot).
Combined LEP limits from direct production 
$e^+ e^- \rightarrow e^\star e$ and from indirect searches
in $e^+ e^- \rightarrow \gamma \gamma $ (virtual $e^\star$ exchange)
are included for comparison.

%

\begin{figure}[h]
\begin{center}
\epsfig{figure=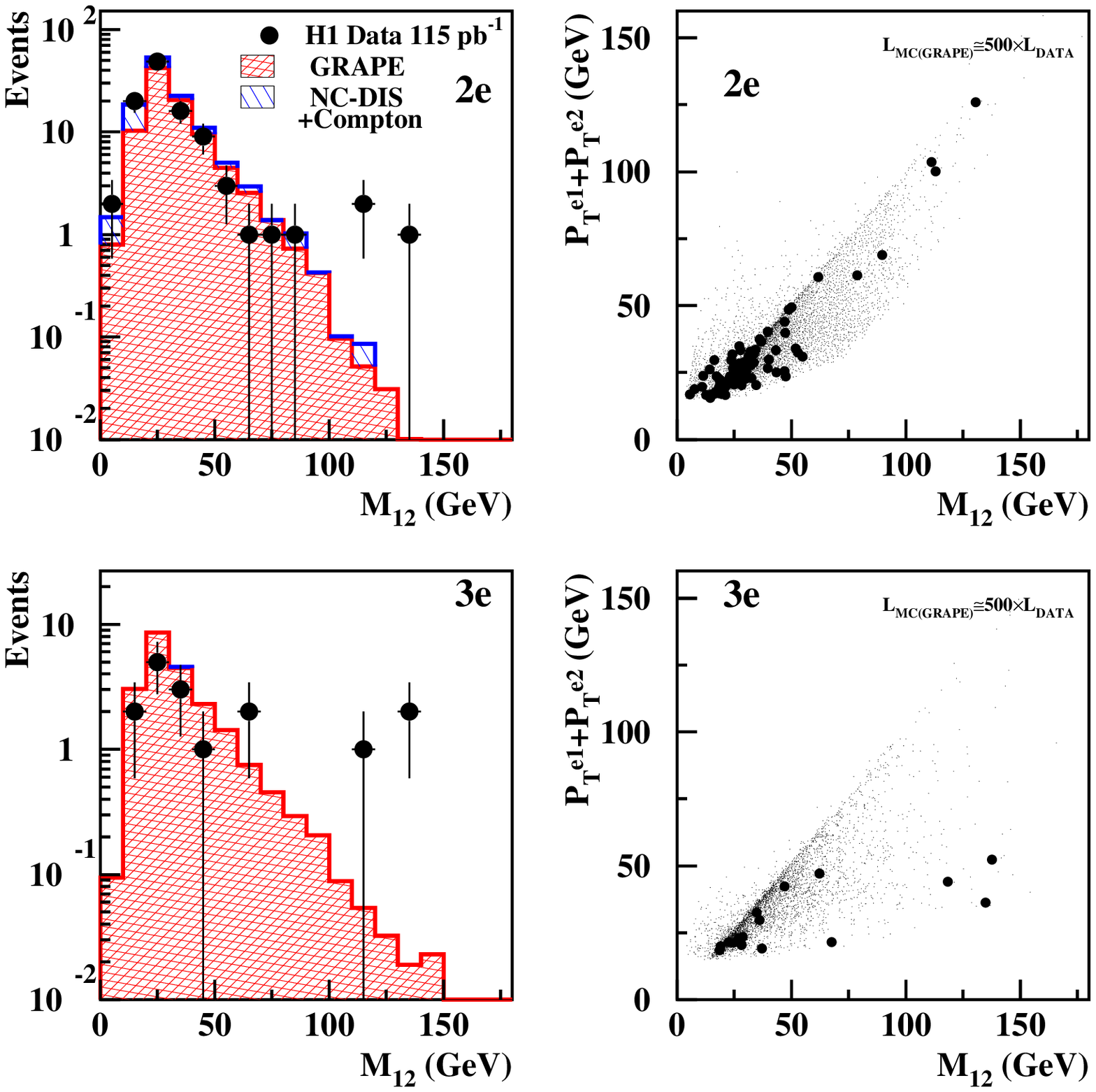,width=6cm,clip=}  
\epsfig{figure=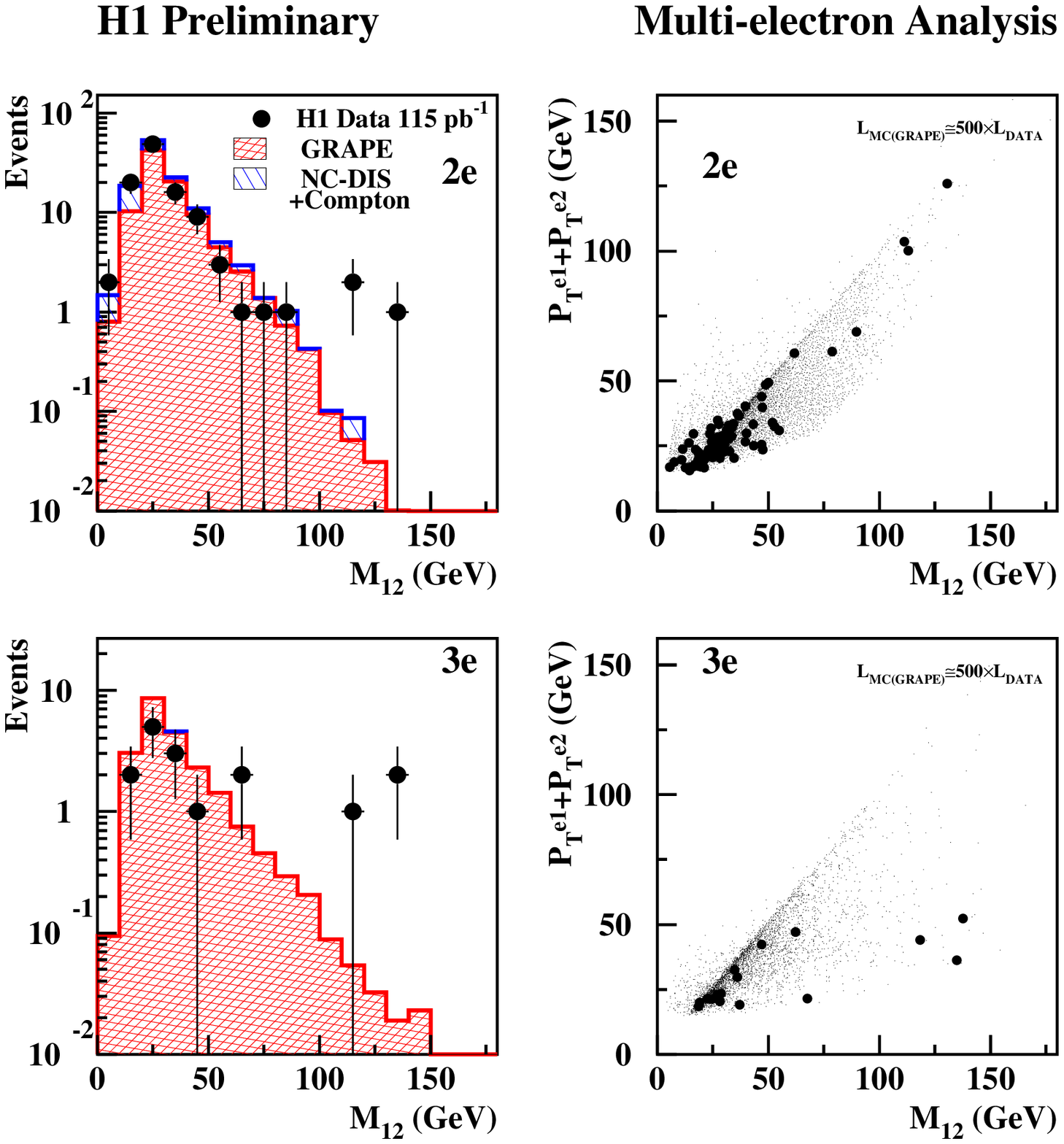,width=6cm,clip=}
\end{center}\vspace*{-.5cm}
  \caption{Distribution of the invariant mass $M_{12}$ of the two
           highest $p_{T}$ electrons, for H1 events classified as
           di-electrons (left) and tri-electrons (right).}
  \label{fig:vallee}
\end{figure}

First measurement of the high-$p_T$ multi-electron production at HERA
was reported by the H1 Collaboration \cite{h1_multie}.
The dominant Standard Model contribution is the interaction of two photons
radiated from the incident electron and proton.
The observed events are in general agreement with Monte Carlo 
predictions.
However, for highest $p_T$ and electron pair invariant masses above 100~GeV
three events classified as di-electrons, and three tri-electrons are 
observed where from the 
SM  only 0.25$\pm$0.05 and 0.23$\pm$0.04 are expected, respectively.
The distributions of the mass $M_{12}$ of the two highest $p_T$ electrons
for di-electron and tri-electron events are shown in Fig.~\ref{fig:vallee}.
This observation, which could be a hint of ``new physics'' beyond SM,
needs confirmation with independent data samples.
\footnote{ZEUS Collaboration presented results on multi-electron
          production at HERA in July 2002 \cite{zeusee}.
          In the combined 1994-2000 data sample two di-electron and
          no tri-electron events are observed, compared to SM
          expectations of $0.77 \pm 0.08$ and $0.37 \pm 0.04$,
          respectively.}

\section{Future prospects}
\label{sec:future}
Both the TeVatron and HERA underwent major upgrades of the machines in
the last several years, and are currently coming online with a new
phase of data taking. At HERA it is now 
also possible to carry on experiments
with polarized lepton beams. In section 6.1 and 6.2 the physics potential 
of these two upgrades is discussed.

In addition we have an outlook on future machines, the LHC and a future LC.
LHC is scheduled to have its first run in 2007, and physics results
will follow within a year or two as the energy frontier will be
extended in the following years.  In particular for searches for
physics beyond the Standard Model the LHC will have great impact
mainly in searches for scalar quarks and gluinos in SUSY. As for
scalar leptons as well as the gaugino/higgsino sector additional help
will be provided by a future Linear Collider (LC) with its first
phase in the energy range of $\sqrt{s}=500$~GeV and upgrade
possibilities to about 1 TeV. Due to the clean environment the LC
will provide measurements with unprecedented precision so that the
underlying structure of the physics can be unambigously revealed.
We summarize in section 6.3 and 6.4 some highlights about the expected 
physics potential of these future machines.

\subsection{HERA upgrade}
The luminosity system of H1 and ZEUS as well as the detectors itself have been
decisively improved \cite{HERA-upgrade}.  The expected integrated
luminosity of HERA II will be 1000~pb$^{-1}$.  Moreover new spin
rotators and polarimeters have been installed at Zeus and H1 so that
also colliding experiments 
with longitudinally polarized $e^{\pm}$ beams will be
done.  The luminosity will be equally shared between $e^+p$ and $e^-
p$ and $L$ and $R$ polarizations. The detectors have both been
improved, e.g. by introducing new or upgraded
micro--vertex detector and improved
triggering. In addition for both experiments the tracking in forward direction
has been upgraded.

\begin{figure}[h]
\begin{center}
  \begin{minipage}{2.3in}\hspace*{-.5cm}\vspace*{.2cm}
\psfig{figure=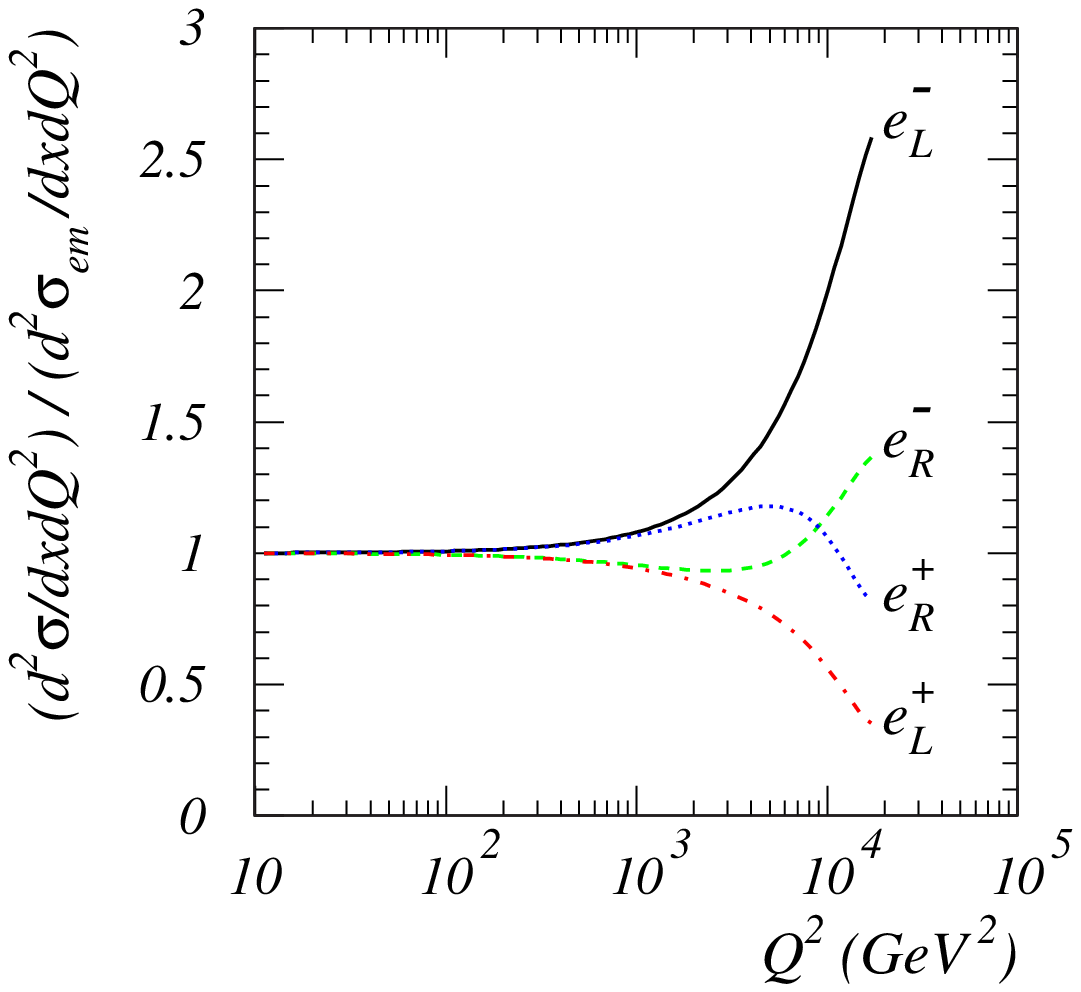,height=2.2in}\vspace*{-.8cm}
\caption{ {\scriptsize  Sensitivity at high Q2 on electroweak effects
         due to difference of cross section for different lepton
         charges and different polarizations.}}
    \label{sens2}
  \end{minipage}\vspace*{-.5cm}
\hfill
  \begin{minipage}{2.3in}
\psfig{figure=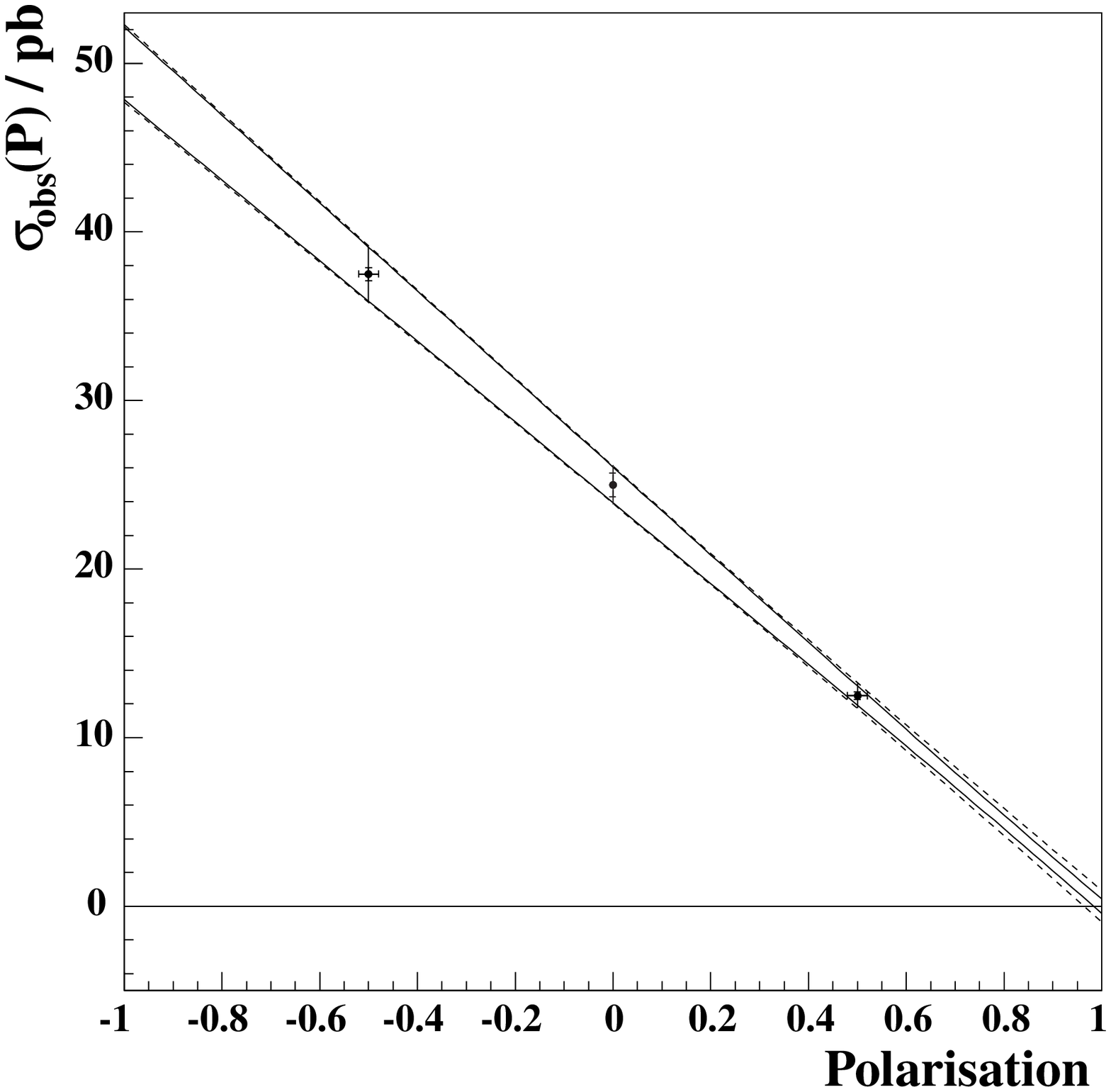,height=1.8in}
\caption{{\scriptsize CC cross section for $e^{-}p$ scattering as a
         function of polarization; P=0: measured point (incl.\ 4.2\%
         systematics).  Errors due to the polarization given by solid
         (dashed) line corresponding to 0.2\% (2\%) polarization
         uncertainty.}}
    \label{ccright} 
  \end{minipage}\vspace*{-.5cm}
\hfill
\end{center}
\end{figure}

Within the SM the NC and CC cross sections are 
affected by the charge and by the longitudinal polarization of the 
incoming lepton. Therefore the use of polarized leptons is very promising.
With different choices of the lepton charge as well as the
polarization one can get
complementary information about 
PDFs as well as the electro--weak couplings in NC
in particular at high $Q^2$~GeV$^2$, see also Fig.~\ref{sens2}.
For the measurements of the $Z^0$ boson couplings to light
quarks $u$, $d$ the beam polarization improves significantly the
accuracy. A run with 250~pb$^{-1}$ will lead to $\delta(a_u)=0.04$,
$\delta(a_d)=0.10$ and with an $e^{\pm}$ polarization of $\ge 50\%$ a
precise extraction of the vector couplings with $\delta(v_u)=0.015$
and $\delta(v_d)=0.04$ is reachable. 
The measurements of HERA are complementary to those of LEP where c--quark 
couplings were probed. 

Furthermore, also for the analysis of the left--handed charged currents
the use of polarized beams will be decisive. Cross sections of charged
currents depend linearly on the beam polarization and any deviation
from this behaviour would be a sign for physics beyond the SM, 
Fig.~\ref{ccright}. 

\subsubsection{Polarization at HERA}

The HERMES fixed gas target experiment used
successfully polarized beams since 1994. In a storage ring the
$e^{\pm}$ beams become `naturally' polarized via the Sokolov--Ternov
effect. In an exactly planar ring, e.g., the (time--dependent)
polarization is vertically orientated. Spin rotators before and after
the intersection region turn the polarization in the longitudinal
direction. Since in reality 
the storage ring is not planar 
depolarizing effects occur and turn the natural maximum polarization of
92.4\% into 50\%--60\% degree which is expected to be reached at
HERA II. For the HERA upgrade two more
regions with spin rotators at H1 and ZEUS 
have been installed but not yet commissioned.
For further details see \cite{Eliana-proc}.

As mentioned before not only the degree of polarization is decisive
but also the accuracy with which this polarization can be measured.
For this purpose significant improvements of the Longitudinal
Polarimeter (LPOL) as well as the Transversal Polarimeter (TPOL) have
been made. The LPOL needs only an energy measurement of the Compton
scattered photon, whereas the TPOL needs to measure its vertical
position in addition and measures then the up--down asymmetry.

At the HERA II upgrade the LPOL is improved by a 
Fabry--Perot cavity in order to run
in a few photon mode which allows for an improved analysis and
leads to high statistics. 
The TPOL on the other hand is
upgraded in a two--fold way. In order to improve the angular
resolution a silicon strip detector has been installed in front of the
calorimeter which allows a continous position calibration during
the measurement.  Furthermore, due to a new data acquisition a bunch to
bunch measurement will be possible so that altogether an accuracy of
$\delta P/P\sim 1\%$ can be reached. Further details can be found in 
\cite{Jenny-proc}.

\subsubsection{Searches at HERA II}

For an integrated luminosity of more than $500$~pb$^{-1}$ 
the upgraded HERA II experiment could make 
a discovery if the H1 effects for  the isolated leptons with $p_T^{miss}$ 
persists in $W$ production.

Also other regions of physics beyond the Standard Model, comparable to
the reachable regions at TeVatron II, are open at the run HERA II. In
particular for the search for scalar leptoquarks the search can be
increased up to a mass scale of about $M_{LQ}\sim 300$~GeV.
But also  in the search for R-parity violating mSUGRA the limits for large 
$\tan\beta$ parameters are competitive to the other experiments.

Very interesting channels are those of  excited leptons, in particular
neutrinos. For low masses $M_{\nu}^*\le 200$~GeV HERA II will be unique to
discover this physics beyond the Standard Model. For higher masses the
DELPHI results will overtake.

Concluding one can state that after the detector commissioning the HERA
II upgrade is complete and has real chance of discovering new physics
beyond the Standard Model. The use of polarized beams will provide an
important tool.

\subsection{TeVatron upgrades}
The Run I data taking period at the TeVatron ended in February 1996. Since 
then the collider and both the detectors (CDF and D0) underwent substantial
upgrades. \par
The energy of the beams has been increased from 900 GeV to 980 GeV. 
A new synchrotron (`` main injector'') has been built in a new tunnel. 
The main injector
together with a debuncher-accumulator-recycler complex allows for 
faster production of 
antiprotons and the possibility of reusing them after they are rescued 
in the recycler.
In Run I the luminosity reached 1.5$\times$10$^{31}$cm$^{-2}$sec$^{-1}$ 
and was 
obtained with a 6 on 6 proton-antiproton bunches in the collider with 
an interbunch 
time of 3.5~${\mu}$sec. The luminosity ultimately  planned for Run II 
is 2.0$\times 10^{32}$cm$^{-2}$sec$^{-1}$
and it 
will be obtained with 36 on 36 proton-antiproton bunches with 
interbunch time of 396 ns.
Eventually, in order to decrease the number of average interactions 
per bunch crossing below 2, the number of bunches in the antiproton beam 
will be increased
to 108 with 140 bunches in the proton beam and a reduced interbunch time of
132 ns.

We have already mention that the quest for Higgs
will be the main topic of research at the TeVatron in the next several years.

For top physics an extra 30-35\% in the cross section is gained (1.8
to 2~TeV).  There will also be a gain from acceptance and efficiency:
100~pb$^{-1}$ in Run II is equivalent to 150-300~pb$^{-1}$ in Run I).
At this time work is still ongoing to finalize b-tagging software
algorithms and the complete understanding of associated background.
Top mass and W mass measurements will be updated from Run I results in
winter 2003. The precision expected on the W mass is of order 20-30
GeV/$c^2$.  The top mass measurement will be improved to a level of
2-3 GeV/$c^2$.  Indirect constraints on the Higgs mass will be of
course derived.  In addition one of the goals of Run II is to search for
$t\bar t$ resonances, rare decays and deviations from the expected
patterns of top decays.  The decay mode where both the W's from top
decay leptonically will be the first to be looked at: in fact a
moderate excess of events in Run I, especially at large missing energy,
is driving the investigation with an eye to signal for new
physics \cite{dilepton}.

\begin{figure}[h]
\begin{center}
  \begin{minipage}{2.3in}\hspace*{-.5cm}
\psfig{figure=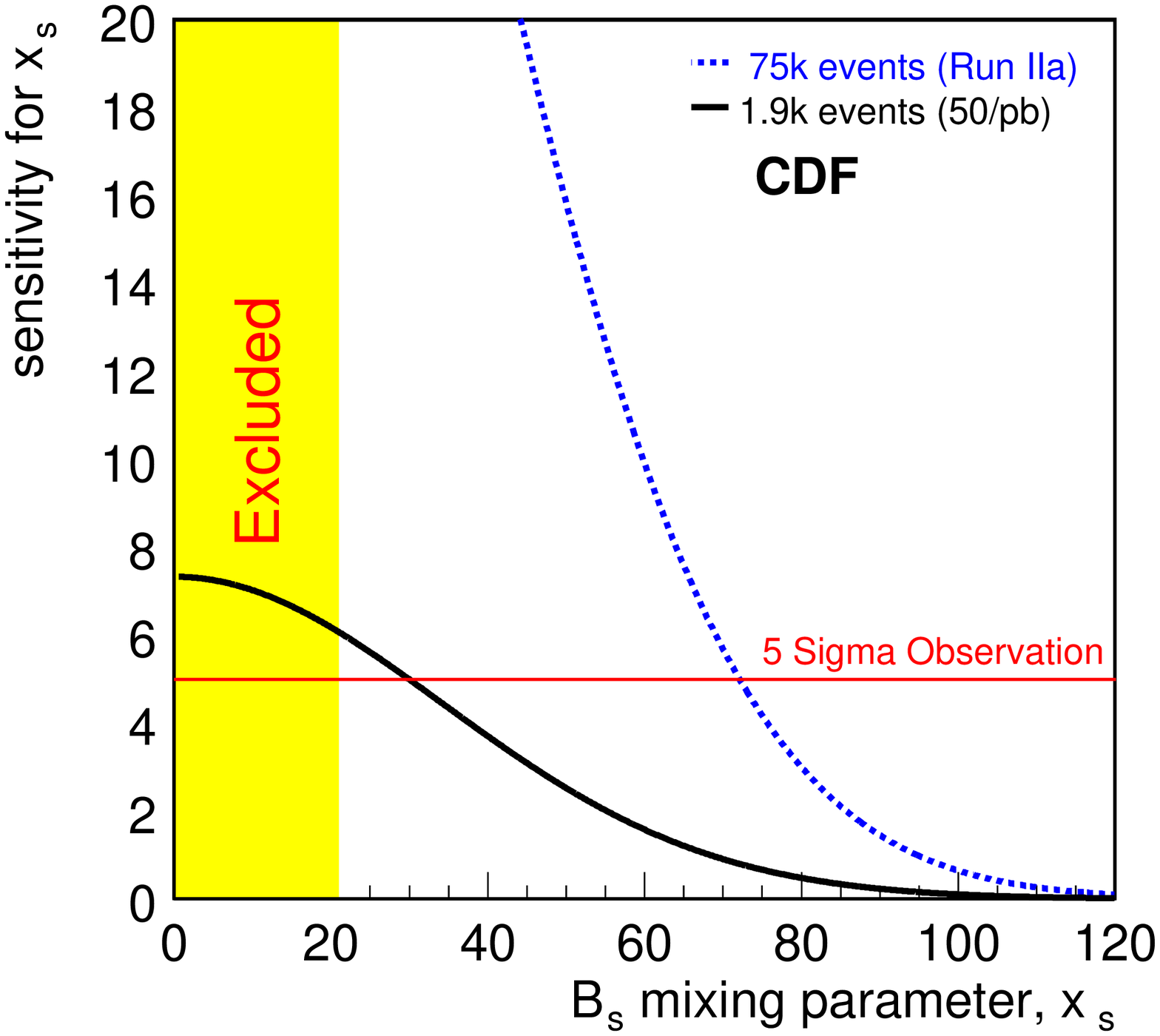,height=1.8in}
\caption{ {\scriptsize  $x_s$ reaches as 
function of number of events expected in 
Run IIa and with 50 pb$^{-1}$}}
    \label{bmixing1}
  \end{minipage}\vspace*{-.3cm}
\hfill\hspace*{-.5cm}
  \begin{minipage}{2.3in}
\psfig{figure=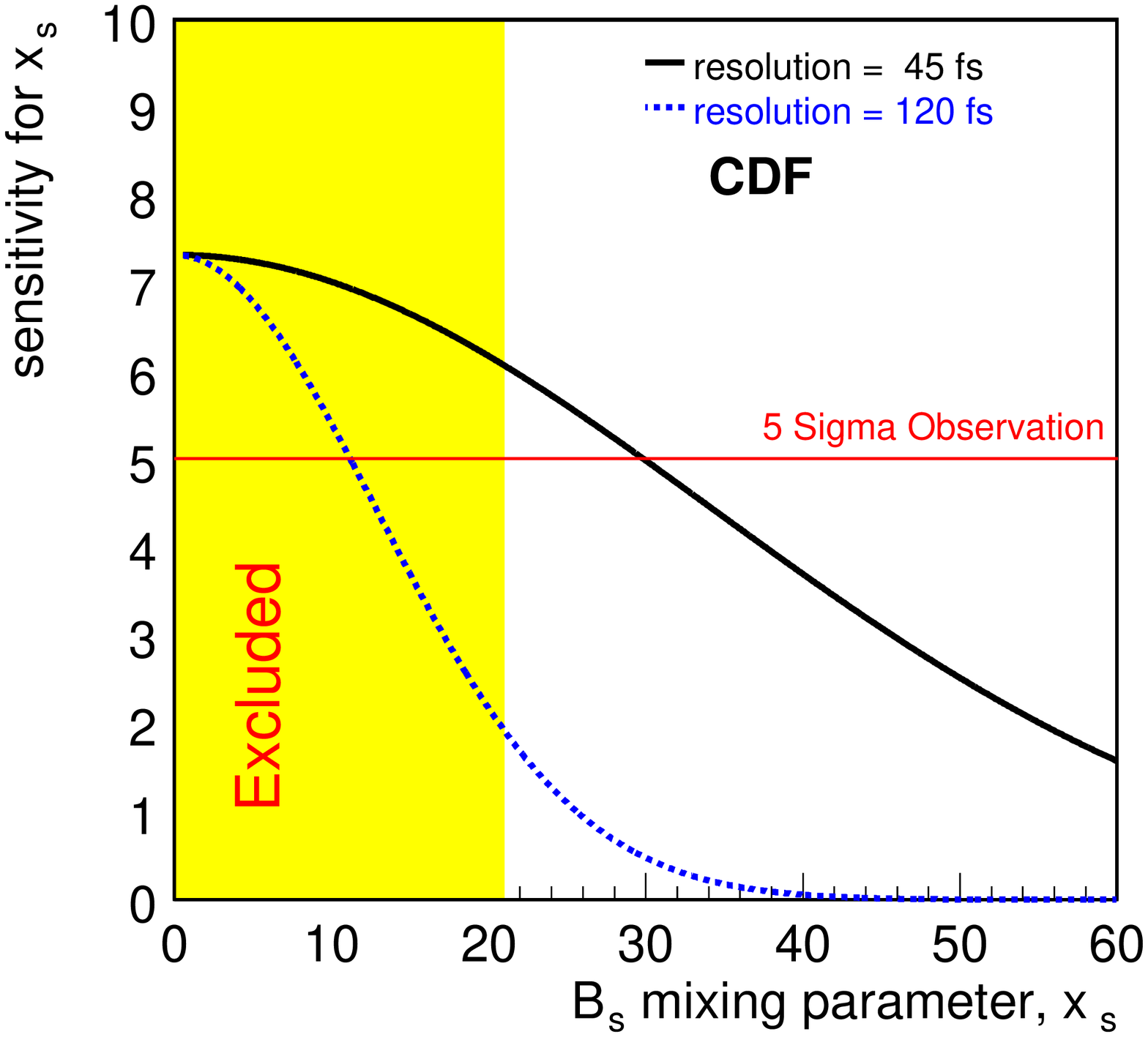,height=1.8in}
\caption{{\scriptsize $x_s$ reaches as function of time re\-solution}}
    \label{bmixing} 
  \end{minipage}\vspace*{-.3cm}
\hfill
\end{center}
\end{figure}

Beauty and charm physics will also receive special attention at the
TeVatron, since the new capabilities of the detectors (possibility of
triggering on displaced vertex tracks) are making the TeVatron
comparable to a dedicated $b/c$ factory.  CP violation in the b sector
is of particular interest as evidence of physics beyond the SM.  In
the framework of the Standard Model, the source of CP violation and B
mixing are the transitions between quarks described by the
Cabibbo-Kobayashi-Maskawa matrix. In this model CP violation arises
due to the irreducible phases in the CKM matrix.  A precision
measurement of the $B_s^0$ flavor oscillations is very important for
testing the unitarity of the CKM matrix~\cite{breport}. The measurement
of $sin 2\beta$, one of the angles of the unitarity triangle, is
obtained by extracting the amplitude of the CP asymmetry in the decay
$B^0/\bar {B^0} \to J/\psi K^0_s$. CDF has measured $sin 2\beta$ with
a precision comparable to that of dedicated B-factories and the
TeVatron Run II will provide the conditions to perform this
measurement with better precision~\cite{wolter}.  The Standard Model
favors a value of the parameter $x_s$ between 22.55 and 34.11 at
95$\%$ C.L.  CDF plans to use the fully reconstructed hadronic $B_s^0$
decays ( $B_s^0 \to D_s^-\pi^+\pi^-\pi^+$ with $D_s^-$ reconstructed
as $\phi \pi^-$, $K^{0*}, K^-$ and $K^-_s,K^-$).  These signals will
come from data taken with the triggers based on displaced-vertex
tracks.  CDF expects 75000 reconstructed $B_s^0$ decays in 2~fb$^{-1}$
using the above decay modes for an estimated signal-to-background
ratio in the range 1:2 to 2:1. In Figs.~\ref{bmixing1} and ~\ref{bmixing} the
expectations for an integrated luminosity of order 50~pb$^{-1}$ are
shown.  The proper time resolution is expected to be in the range
45-60 fs 
and the flavor tag effectiveness ($\epsilon D^2$) around
11\%.  This value includes same-side tagging, soft lepton tagging and
opposite-side jet tagging, as well as kaon tagging now made possible
by the use of the TOF detector.

The expected total integrated luminosity for Run II will allow to search
more efficiently for physics beyond the Standard Model.
CDF will search for SUSY particles in first place.
Assuming that SUSY breaking results in universal soft breaking 
parameters at the grand unification scale, and that the lightest 
supersymmetric particle is stable and neutral, 
with 30 fb$^{-1}$ luminosity and one detector, 
charginos and neutralinos, as well as third generation squarks, 
can be seen if their masses are not larger than 200-250 GeV, 
while first and second generation squarks and gluinos can be discovered 
if their masses do not significantly exceed 400 GeV~\cite{sugra}.

\begin{figure}[h]
\begin{center}
\psfig{figure=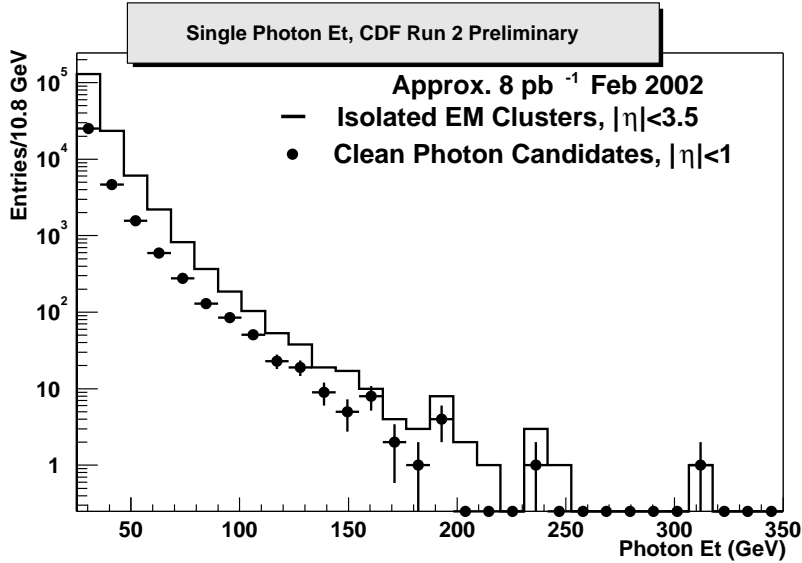,height=3.0in}
 \caption{\it
     First photon candidate events from CDF Run II. 
    \label{figure4} }
\end{center}\vspace*{-1cm}
\end{figure}

Models where  SUSY is broken at low scale including 
gauge-mediated supersymmetry breaking are generally distinguished by the 
presence of a nearly massless Goldstino as the lightest supersymmetric 
particle. 
The next-lightest supersymmetric particle(s) (NLSP) decays to its 
partner and the Goldstino. Depending on the supersymmetry breaking scale, 
these decays can occur promptly or on a scale comparable to or larger 
than the size of a detector. A systematic analysis based on a 
classification in terms of the identity of the NLSP and its decay length 
has been presented for example in
\cite{gmsb}. The various scenarios have been  discussed in terms of 
signatures and 
possible event selection criteria. Analysis are starting in CDF with the
aim of understanding the datasets in terms of background contribution and
possible deviation from it as a sign of new physics. 
Signatures involving photons are of particular interest to look for
deviations from the SM predictions in the context of GMSB models.
CDF is also using photon signatures as a first follow-up and check of
strange events seen in Run I: in Fig.~\ref{figure4} 
the spectra of single photons
candidate is reported using approximately 8~pb$^{-1}$ of Run II data.

\subsection{LHC}
\begin{figure}[h]
\begin{center}
\epsfig{figure=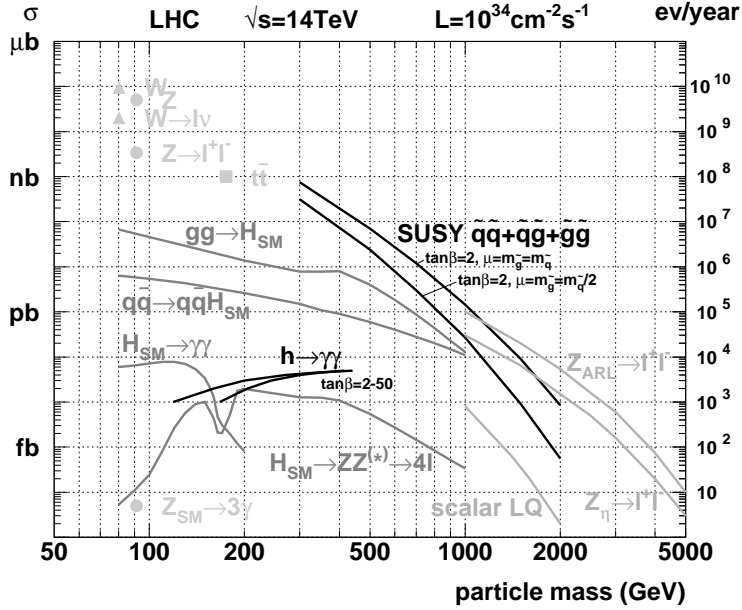,width=12cm,clip=}
\end{center}\vspace*{-1cm}
  \caption{Cross sections and event rates (per one year of
           LHC running at nominal luminosity) for different
           SM and beyond-SM processes which can be studied at the LHC.}
  \label{fig:lhc}\vspace*{-.3cm}
\end{figure}

The LHC will offer a large range of physics opportunities due to the
high energy, $\sqrt{s}$ = 14~TeV, and a high luminosity up to ${\cal
L} = 10^{34}$~cm$^{-2}$~s$^{-1}$.
The first physics run is foreseen to start in 2007.
The cross sections and expected production rates 
of many relevant physics processes are large, as can be seen
in Fig.~\ref{fig:lhc}.
Event samples, 
which will be collected at the LHC,
will allow to perform many precision measurements 
of the Standard Model and possible beyond-SM scenarios \cite{lhc_phys}.

Large $t\bar{t}$ samples will allow to measure  
the top quark mass up to 1--2 GeV,
the production cross section of about 5\%
and also provide the detailed study of 
branching ratios, couplings and 
rare decays such as the flavour--changing neutral current reactions down to
branching ratios of about $10^{-4}$.
The SM Higgs boson production cross section 
is larger than 100~fb up to 1~TeV,
and the discovery is possible over the full mass range 
from the LEP2 lower limit up to the TeV range, already with 10 fb$^{-1}$.
After few years of LHC running with high luminosity
ATLAS+CMS would 
measure the Higgs boson mass and
width, as well as production rates (with a precision of $\sim$10\%),
and moderate measurements of some 
couplings 
with certain model--assumptions 
($\sim$10-25\%).

Searches beyond the Standard Model are among 
the important goals of the ATLAS and CMS experiments.
LHC has a large discovery potential for Higgs bosons in
the MSSM Higgs sector, 
two or more Higgs bosons should be observable 
over large portions of parameter space. 
However, there remains a region at $m_A >  200$ GeV 
and $4 \le  \tan \beta \le 10$ where only the $h^\circ$ would be seen
and it would be indistinguishable from a SM Higgs.

Excited quarks and leptons should be observable in different channels 
up to masses of about 7~TeV.
New gauge bosons can be discovered up to masses of 5~TeV.
The existence of right-handed W or heavy Majorana neutrinos 
can be assessed up to a few TeV, 
and charged heavy leptons can be discovered up to masses of about 1.1~TeV.

In theories of large extra dimensions the fundamental Planck scale 
can be of the order of TeV.
Large effects from real graviton production or  virtual graviton 
exchange are expected at the LHC.
Jets or photons in conjunction with missing transverse energy are 
considered as a signature for graviton emission.
A signal will be observable at LHC if the Planck scale 
in $4+\delta$ dimensions is below 9~TeV for $\delta = 2$,
or 6~TeV if $\delta = 4$.
The decay mode $G \rightarrow l^+ l^-$   gives a good signal 
of narrow graviton resonances up to 2.1~TeV, 
if the Randall-Sundrum scenario is used.
The presence of the virtual graviton exchange contribution
to Drell-Yan processes can lead to a significant excess 
in the production of dilepton and diphoton events, if
the fundamental Planck scale is below 8~TeV.

If the fundamental Planck scale is below few TeV,
LHC could then turn into a black hole factory.   
The non-perturbative process of black hole formation and
decay by Hawking evaporation gives rise to spectacular events with up to
many dozens of relatively hard jets and leptons~\cite{bh}.
For production of black holes more massive
than 5~TeV at the LHC with $M_p=1$~TeV and $\delta=10$,
the integrated cross section function would be of the order of $10^5$~fb,
corresponding to a production rate of a few Hz.
Even for black holes more massive than 10~TeV 
a production rate of a few per day might be expected.
With TeV scale gravity, black hole production could become the dominant
process at hadron colliders beyond the LHC.

\subsection{Linear Collider}

An $e^+e^-$ Linear Collider with $\sqrt{s}=500\ldots 1000$~GeV has
been recognized to be the next major machine to be built for high energy physics research.  
It offers the possibility of a very 
precise analysis of the physics at the TeV
scale. In this context the recently formed LHC / LC Study Group
\cite{LHCLC}, which works out the optimal path for a hand--in--hand
research of the hadron machines and the LC, has already revealed a
large number of topics where using the results of one machine as input
for the analysis at the other machine can be very fruitful.

The use of polarized beams \cite{polarisation}, tunable center of mass energy
as well as the different options
$e^+ e^-$, $e^- e^-$, $e^{\pm}\gamma$, $\gamma \gamma$ \cite{TDR} 
offers a high degree of flexibility.

\subsubsection{Top physics}

Due to the high precision reachable at the LC,
the mass and couplings of the top quark will be improved by at least
one order of magnitude w.r.t. the LHC.  At a threshold scan with an integrated
luminosity of about $100$~fb$^{-1}$ and both beams polarized 
($|P_{e^-}|=80\%$, $|P_{e^+}|=60\%$) the mass will be measured with an
accuracy of about $\delta(m_t)=100$~MeV, Fig.~\ref{lctop}, and
$\delta(\Gamma_t)/\Gamma_t=0.05$.  The vector coupling to a relative
precision of 2\% (or even 0.8\% with $300$~fb$^{-1}$) \cite{TDR}
will become sensitive to quantum corrections.

\subsubsection{Higgs physics}

\begin{figure}[h]
\begin{center}\vspace*{-.5cm}
  \begin{minipage}{2.3in}\hspace*{-.6cm}\vspace*{-.3cm}
\psfig{figure=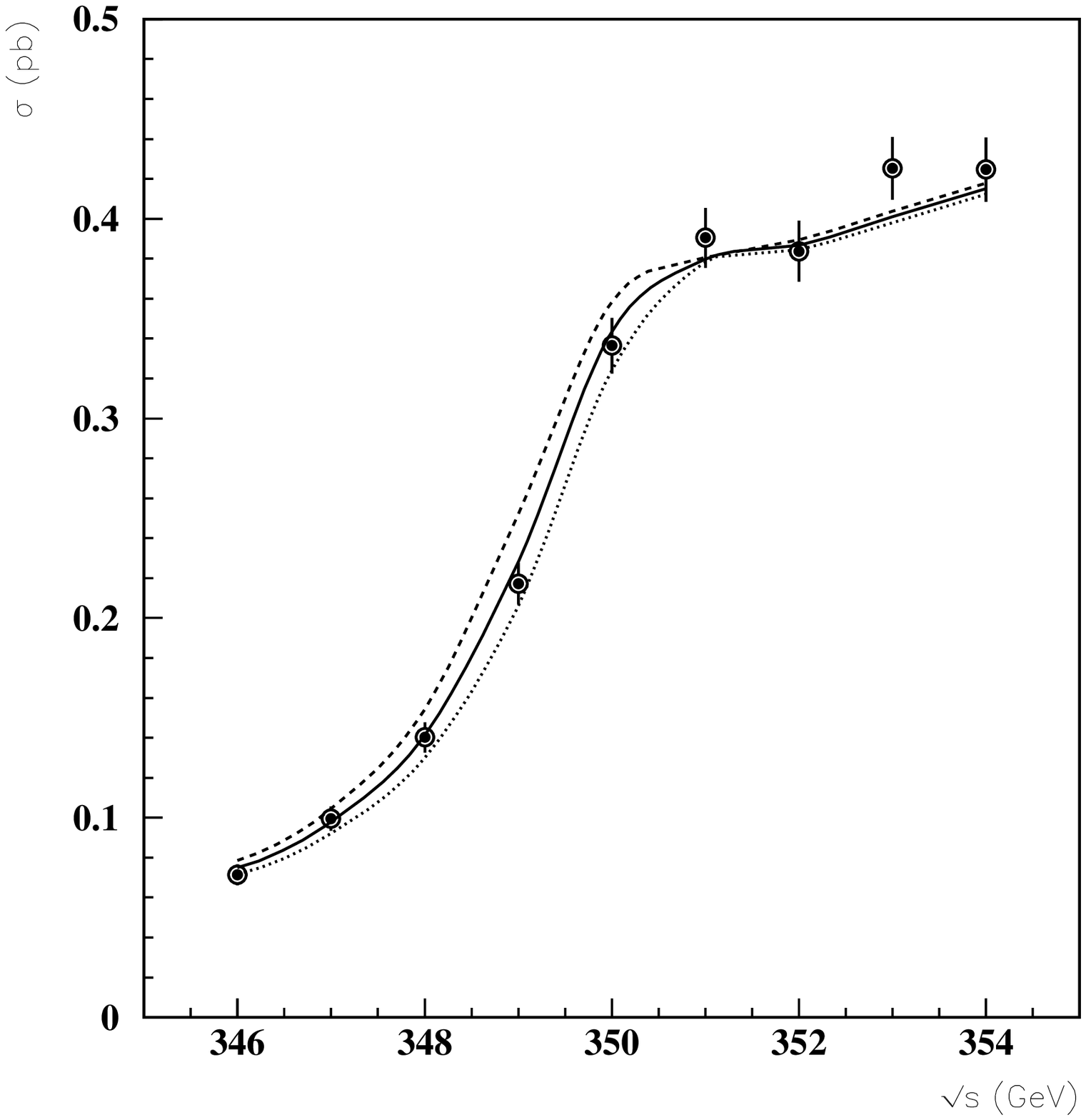,height=2.7in}\vspace*{-.2cm}
\caption{ {\scriptsize Excitation curve of $t\bar{t}$ (including errors bars 
for $100$~fb$^{-1}$). The dotted curves indicate shifts of the top mass by
$\pm 100$~MeV. }}
    \label{lctop}
  \end{minipage}\vspace*{.6cm}
\hfill
\begin{minipage}{2.3in}\hspace*{-.2cm}
\psfig{figure=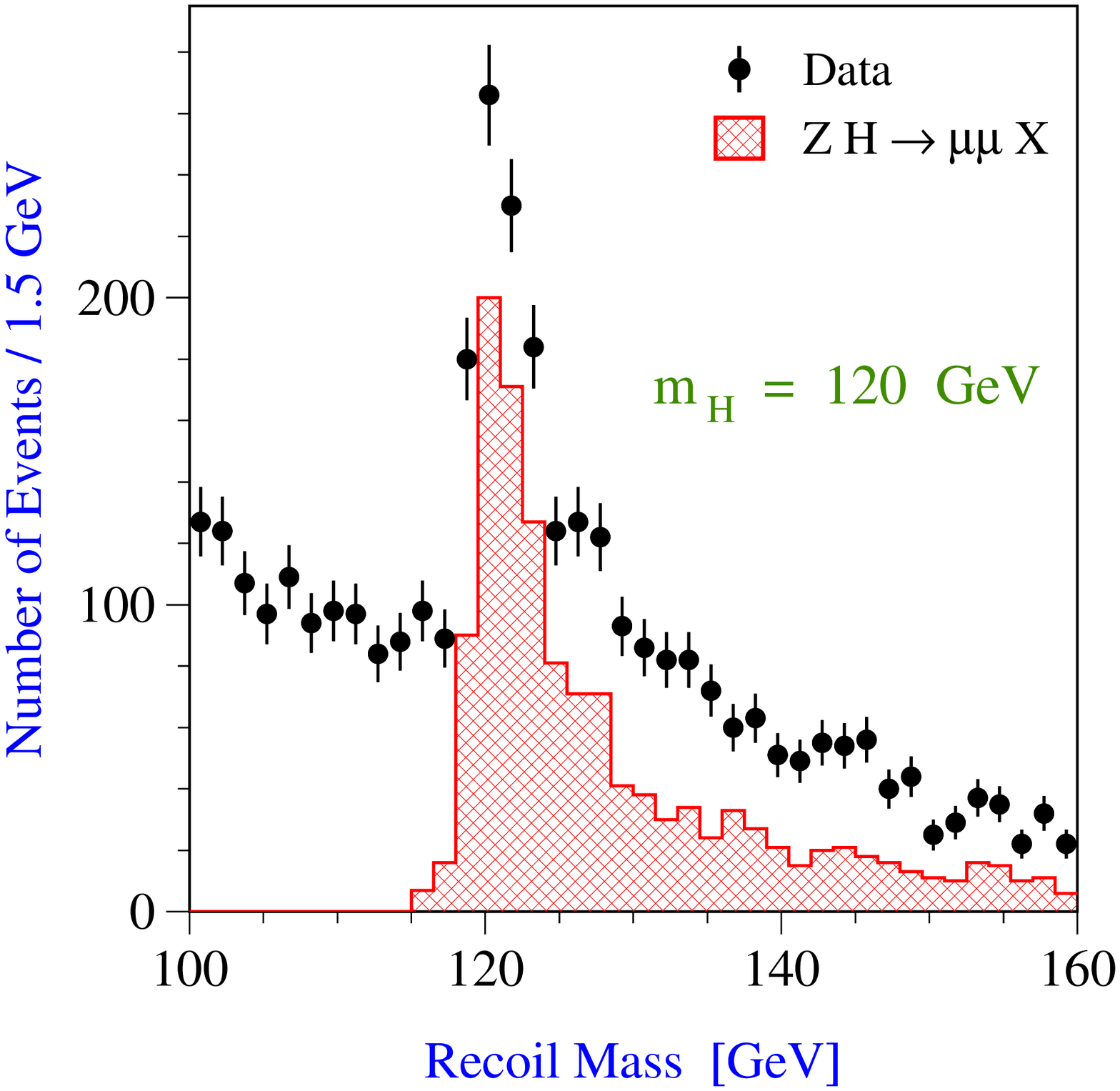,height=2.0in}
\caption{ {\scriptsize The $\mu^+\mu^-$ recoil mass distribution in
the process $e^+e^-\to H Z\to X \mu^+\mu^-$ for $M_h=120$~GeV and
500$^{-1}$ at $\sqrt{s}=350$~GeV. The dotted error bars are Monte
Carlo simulation of Higgs signal and background. The shaded histogram
represents the signal only.}} \label{lchiggs}
\end{minipage}\vspace*{-.5cm}
\hfill
\end{center}
\end{figure}

The properties of the Higgs bosons can be determined with high
precision and all essential elements of the mechanism of electroweak
symmetry breaking can be established. The 
model independent determination of the Higgs mass, Fig.~\ref{lchiggs}, 
the total width and the measurement of
all relevant couplings to bosons and fermions can be performed at the
per cent level. The quantum numbers ($J^{CP}$) can be uniquely
determined. Even all self--coupling of the Higgs boson, which proves
the shape of the Higgs potential, can be measured.

A very interesting option is the use of the $e^{\pm}\gamma$ and
the $\gamma\gamma$ beams at a LC where in particular the
production of the Higgs bosons in the s--channel is possible, 
Fig.~\ref{lcgamma}.

\subsubsection{Anomalous gauge couplings}

\begin{figure}[h]
\begin{center}
  \begin{minipage}{2.3in}\hspace*{-.2cm}
\psfig{figure=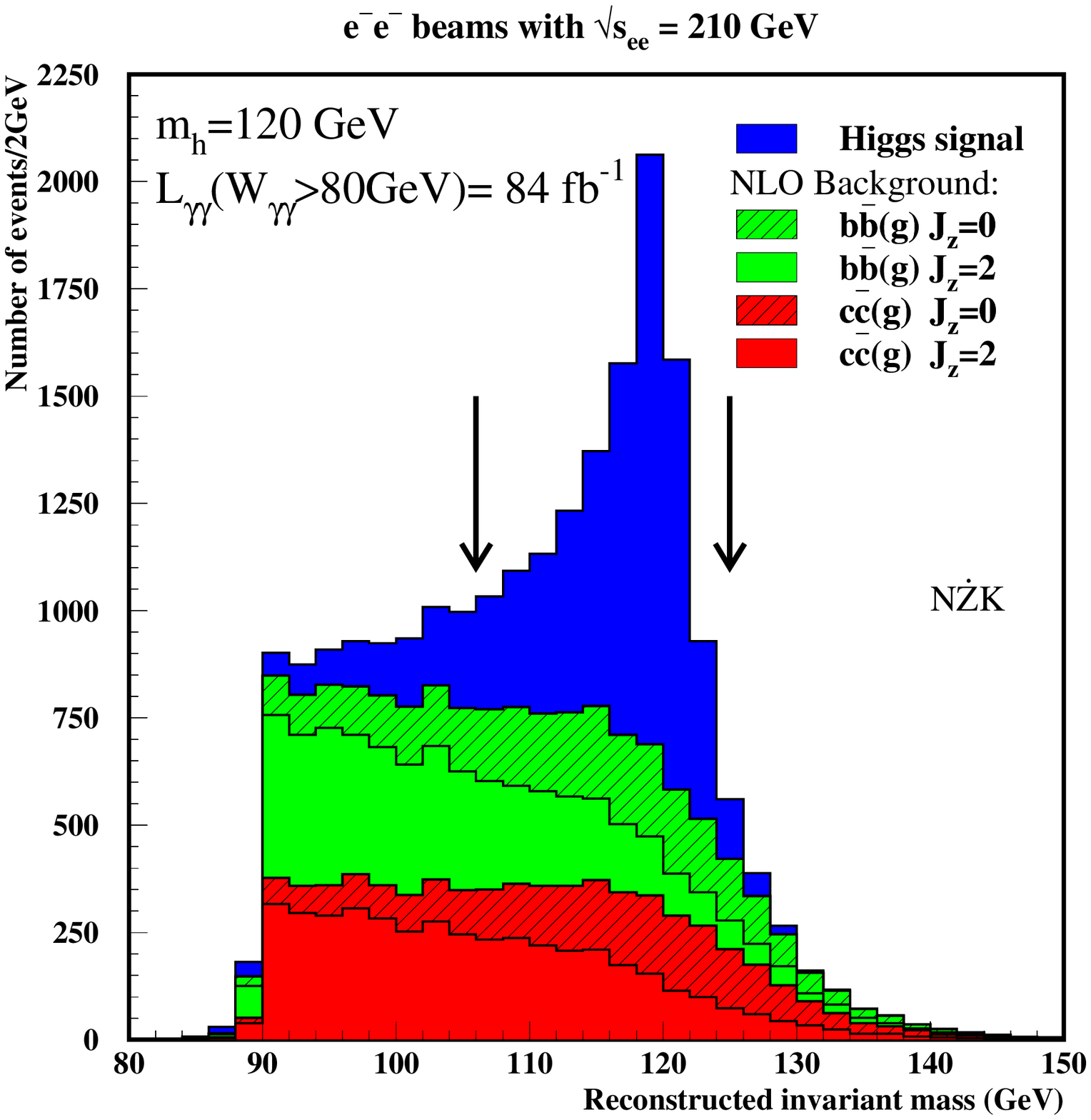,height=2.4in}\vspace*{-.4cm}
\caption{{\scriptsize Distribution of the reconstructed mass 
           for Higgs boson production in 
           $\gamma \gamma$ interactions
           at TESLA, for $m_h$=120 GeV \cite{Filip}. }}
    \label{lcgamma} 
  \end{minipage}\vspace*{-.3cm} \hfill
\hspace*{-.5cm}
  \begin{minipage}{2.3in}\vspace*{-1cm}
\psfig{figure=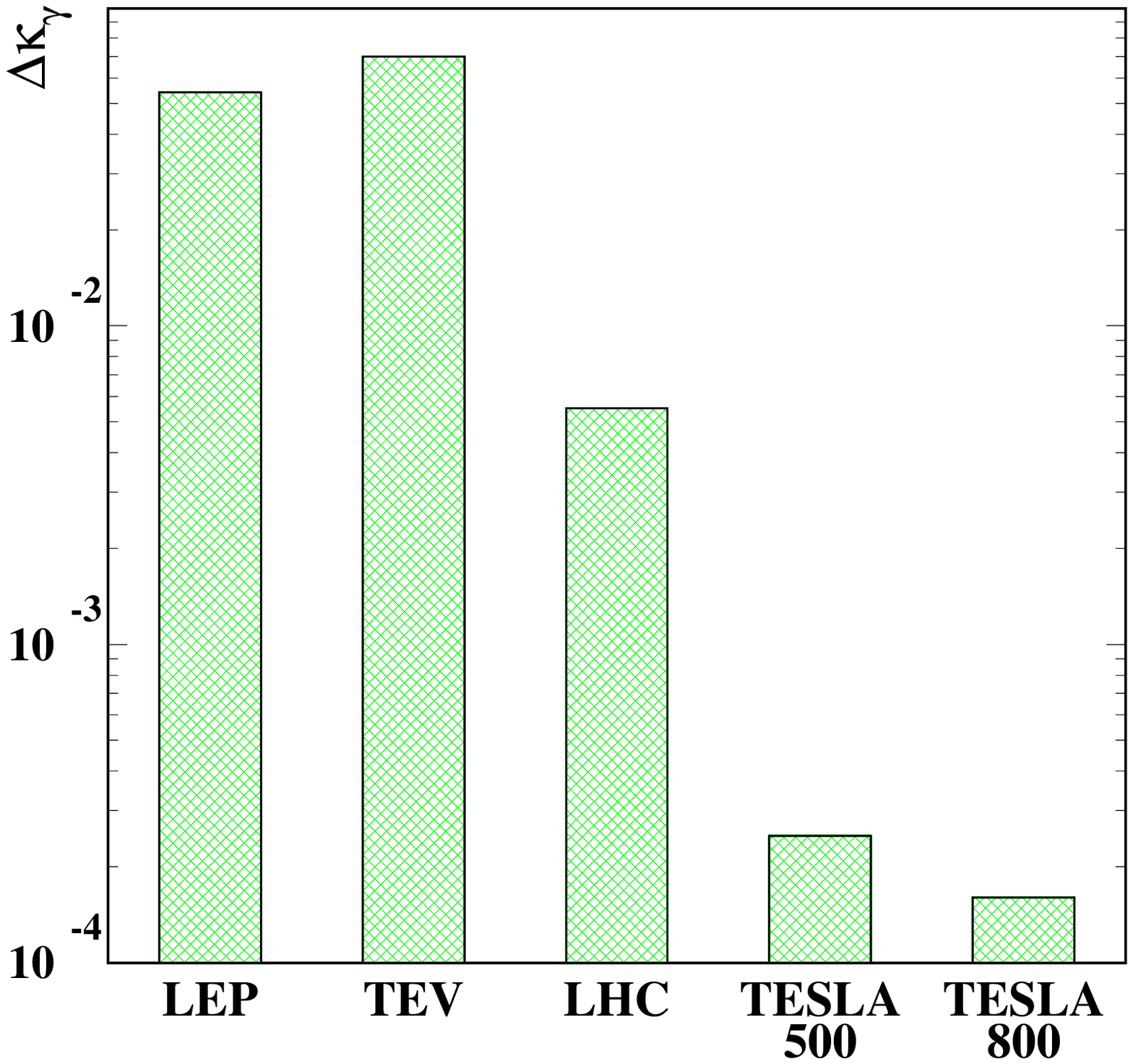,height=2.3in}
\caption{{\scriptsize The precision for anomalous gauge couplings in 
comparison at different machines. For LHC and TESLA three years of running
are assumed. (LHC: 300~fb$^{-1}$, TESLA $\sqrt{s}=500$~GeV: 900~fb$^{-1}$,
TESLA $\sqrt{s}=800$~GeV: 1500~fb$^{-1}$).}}
    \label{lckappa} 
  \end{minipage}\vspace*{-.3cm}
\hfill
\end{center}
\end{figure}

Triple gauge couplings can be measured with an superior accuracy
at the LC providing a high sensitivity to any kind of new physics effects.
In Fig.~\ref{lckappa} the accuracy for triple gauge couplings
at different LC energies are compared to the TeVatron and LHC potential.

\subsubsection{Physics beyond the SM}

As for physics beyond the SM, the LC has a large discovery potential
for supersymmetric particles, in particular for the non--coloured
particles. Due to its clean signatures and low background processes, the LC
can shed light on the structure of the underlying theory very
precisely. 

Since SUSY is not an exact symmetry, but has to be broken at
low--energy, there are 105 new SUSY parameters in addition to the free
19 SM parameters.  Therefore the LC is a unique tool for revealing
the underlying structure of the model and has the challenging task  of the
precise determination of these parameters. 
Experimental and theoretical strategies have been worked
out to determine precisely the low--energy electro weak parameters and
after determining these free parameters powerful consistency tests are
possible in order to understand the SUSY breaking scheme.

A lot of studies for other kinds of physics beyond the SM as phenomena of
R--parity violating SUSY, large extra dimension, extended gauge boson 
sectors have been made and demonstrate the rich physics program of a LC
 \cite{TDR}.

\section{Conclusions}
It has been  a very interesting session and
we look forward to new data from HERA II and the TeVatron!
Let's thank all our speakers:
J. Boehme, S. Chivucula, N. De Filippis, P. Deglon, K. Desch,
M. Ellerbrock, C. Foudas, C. Genta, E. Gianfelice,
S. Grijpink, J. Haller, M. Helbich, J. Kalinowski, M. Krawczyk, Z. Lalak,
N. Malden, S. Mattingly, A. Mehta, 
C. Mesropian, S. Moch, 
M. Moritz, M. Petteni, K. Piotrzkowski,
T. Pratt, T. Saeki, S. Schmidt, K. Sliwa, 
P. Spentzouris, R. Stroehmer, J. Sztuk,
C. Vallee, A. Weber, G. Weiglein, M. Wolter, G. Wrochna, P. Zalewski.

\section*{Acknowledgements}

The authors thank Jan Kwiecinski and the other organizers of DIS 2002 
for the extremely hostly and friendly atmosphere during the Conference.
GMP thanks J.~B\"ohme, K.~Desch, 
E.~Elsen, H.-C.~Schultz-Coulon, H.~Spiesberger and G.~Weiglein 
for interesting discussions. GMP was partially supported by
the Graduiertenkolleg `Zuk\"unftige Entwicklungen in der Teilchenphysik' of
the University of Hamburg, Project No. GRK 602/1. AFZ was partially 
supported by the Polish State 
Committee for Scientific Research (KBN) grant No. 2 P03B 070 22.



\begin{thebibliography}{100}

\bibitem{saeki} T. Saeki, {\it Acta Phys. Pol.} {\bf B33}, 3831 (2002),
these proceedings.

\bibitem{Weiglein} G. Weiglein, Talk at DIS 2002, see also
M.~Carena, S.~Heinemeyer, C.E.M.~Wagner and G.~Weiglein, hep-ph/9912223;
S.~Heinemeyer, W.~Hollik and G.~Weiglein, Eur.\ Phys.\ J.\ C9, 343 (1999);
M.~Frank, S.~Heinemeyer, W.~Hollik and G.~Weiglein, hep-ph/0202166;
A.~Freitas, W.~Hollik, W.~Walter and G.~Weiglein,
Phys. Lett. B495, 338 (2000); Nucl. Phys. B632, 189 (2002).

\bibitem{panagiotis} P.~Spentzouris, {\it Acta Phys. Pol.} {\bf B33}, 3843 
(2002), these proceedings.

\bibitem{foudas} C. Foudas, talk at this conference not submitted to the 
proceedings.

\bibitem{christina} C. Mesropian, {\it Acta Phys. Pol.} {\bf B33}, 3135 (2002),
these proceedings,
FERMILAB-CONF-02-136-E,CDF-PUB-JET-PUBLIC-6008.

\bibitem{chris} K. Sliwa, {\it Acta Phys. Pol.} {\bf B33}, 3861 (2002),
these proceedings.


\bibitem{Mumbai} K. Sliwa ``Top mass and cross section results from CDF and 
D0 at the Fermilab Tevatron", 13th Topical Conference on Hadron 
Collider Physics, 
Tata Institute of Fundamental Research, Mumbai, India, January 14-20, 1999;
in Proceedings, p 169; World Scientific, 1999.

\bibitem{higgs2} LEP ElectroWeak Working Group,\\
http://lephiggs.web.cern.ch/LEPHIGGS/papers/index.html.

\bibitem{petteni} M. Petteni, {\it Acta Phys. Pol} {\bf B33}, 3855 (2002),
these proceedings.

\bibitem{snowmass} T.Munar, S. Rolli, 
presentation given at the Snowmass conference on the future of particle 
physics, Snowmass July 2001, 
http://ncdf70.fnal.gov:8001/talks/snowmass/svt/index.htm .

\bibitem{fnal_runII} M. Carena, J. S. Conway, H. E. Haber, 
J. D. Hobbs, et al, Fermilab-Conf-00/279-T and SCIPP-00/37.


\bibitem{lep_susy_h}
LEP2 Higgs Working Group, ALEPH, DELPHI, L3, OPAL Experiments,
LHWG-2001-04.
\bibitem{carena} M. Carena, S. Heinemeyer, C.E.M.~Wagner, G. Weiglein,
                 hep-ph/9912223.
\bibitem{lep_susy_part}
LEP2 SUSY Working Group, ALEPH, DELPHI, L3, OPAL Experiments,
LEPSUSYWG/02-01.1, LEPSUSYWG/02-02.1, 
see web page \textit{http://lepsusy.web.cern.ch/lepsusy/}.
\bibitem{cdf_susy_new}
CDF Collaboration, T. Affolder et al., Phys. Rev. Lett. 88, 041801 (2002);
D0 Collaboration, B. Abbott et al.,   Phys. Rev. Lett. 83, 4937 (1999).
\bibitem{d0_stop} 
D0 Collaboration, V. M. Abazov et al., Phys. Rev. Lett. 88, 171802 (2002).
\bibitem{hera_mssm}
ZEUS Collaboration; J.Breitweg et al. 
Phys. Letters B 434 (1998) 214-230.
\bibitem{hera_rp}
ZEUS Collaboration
\textit{Search for squark production in R-parity violating 
  interactions at HERA},
Contributed paper \#1042 to the XXXth International Conference 
 on High Energy Physics ``ICHEP 2002'' (July 2000).



\bibitem{add}
N.~Arkani-Hamed, S.~Dimopoulos and G.~Dvali,
                {\it Phys. Lett.} {\bf B429}, 263 (1998);
%
I.~Antoniadis, N.~Arkani-Hamed, S.~Dimopoulos and G.~Dvali,
                {\it Phys. Lett.} {\bf B436}, 257 (1998);
%
N.~Arkani-Hamed, S.~Dimopoulos and G.~Dvali,
                {\it Phys. Rev.} {\bf D 59}, 086004 (1999).


\bibitem{Randall:1999ee}
L.~Randall and R.~Sundrum,
``A large mass hierarchy from a small extra dimension,''
Phys.\ Rev.\ Lett.\  {\bf 83}, 3370 (1999) [hep-ph/9905221].


\bibitem{lep_ed_dir} 

ALEPH Collaboration, 
 ``Single- and Multi-Photon Production and a Search for Slepton Pair 
 Production in GMSB Topologies in $\rm e^+e^-$ Collisions at $\sqrt{s}$ 
 up to 208~GeV'', 
ALEPH-2001-010, CONF-2001-007, 21 February 2001;

DELPHI Collaboration, 
``Update at 202-209~GeV of the Analysis of Photon Events with Missing Energy'',
DELPHI-2001-082, CONF-510, 15 June 2001;

L3 Collaboration, M.~Acciarri et al., Phys.~Lett.~B470 (1999) 268;

OPAL Collaboration, G.~Abbiendi et al., Eur.~Phys.~J. C18 (2001) 253.

\bibitem{tev_ed_dir}

CDF Collaboration, D.Acosta et al., hep-ex/0205057,
 submitted to Phys. Rev. Lett.

\bibitem{add_lep_pred}

G.F.~Giudice, R.~Rattazzi and J.D.~Wells,
Nucl. Phys.{} {\bf B544},~3~(1999).

\bibitem{lep_ed_vir}  

D. Bourilkov
hep-ex/0103039;

OPAL Collaboration, 
``Limits on Low Scale Quantum Gravity in Extra Spatial Dimensions 
from Measurements of $\rm e^+e^- \rightarrow e^+e^-$ at LEP2'', 
OPAL Physics Note 471, 23 February 2001;

DELPHI Collaboration,
''Determination of the $\rm e^+e^- \rightarrow \gamma\gamma(\gamma)$ 
Cross-section Using Data Collected with the DELPHI Detector 
up to the Year 2000'',
DELPHI 2001-093, CONF-521, 1 June 2001;

L3 Collaboration,
``Hard-Photon Production at $\sqrt{s}=192-208$~GeV'', 
L3 Note 2687, 2 July 2001.


\bibitem{tev_ed_vir}

DO Collaboration, D.~Abbott et al.,
Phys.\ Rev.\ Lett. {\bf 86},~1156~(2001).



\bibitem{hera_ci}
H1 Collab., C. Adloff et al., Phys. Lett. B479 (2000) 358-370;
``Search for Compositeness, Leptoquarks and Large Extra Dimensions 
  in eq Contact Interaction at HERA'';\\
H1 Collab., \textit{Search for new physics phenomena in ep contact
                            interactions at HERA },
H1prelim-02-062, contributed paper \#979 
to International Conference on High Energy Physics ``ICHEP 2002'' (July 2002);


ZEUS Collaboration; J.Breitweg et al. 
The European Physical Journal C 14 (2000) 2, 239-254;\\
ZEUS Collaboration, Contributed paper \#602 to the Europhysics Conference on
             High Energy Physics "EPS 2001" (July 2001).

\bibitem{brw} W.~Buchmuller, R.~Ruckl and D.~Wyler,
    Phys.\ Lett.\ B {\bf 191} (1987) 442, 
[Erratum-ibid.\ B {\bf 448} (1999) 320].


\bibitem{hera_lq}

 H1 Collab., C.~Adloff et al.,  Eur.~Phys.~J. C11:447, 1999
              ({\it Erratum} Eur.~Phys.~J. C14:553, 2000);\\
 H1 Collab., C.~Adloff et al., Phys. Rev. Lett. 369:173, 1996;\\
 H1 Collab., C.~Adloff et al., Phys. Lett. B523:234, 2001;\\
H1 Collab., \textit{A search for leptoquarks at HERA},
H1prelim-02-064, contributed paper \#1027 
to International Conference on High Energy Physics ``ICHEP 2002'' (July 2002); 


ZEUS Collaboration; J. Breitweg et al., 
   Physical Review D 63 (2001) 052002;\\
ZEUS Collaboration; J. Breitweg et al.,
       The European Physical Journal C 16 (2000) 2, 253-267;\\
ZEUS Collab., \textit{Search for leptoquarks in ep collisions at HERA},
contributed paper \#907 
to International Conference on High Energy Physics ``ICHEP 2002'' (July 2002). 


\bibitem{tev_lq}

D0 Collaboration, B. Abbott et al.,
Phys. Rev. Letters {79} 4321 (1997),
Phys. Rev. Letters {80} 2051 (1998),
Phys. Rev Lett. {81} 38 (1998),
Phys. Rev. Lett. {83}, 2896 (1999),
Phys. Rev. Lett. {84}, 2088 (2000).


CDF Collaboration, F. Abe et al., 
Phys. Rev. Lett. 79, 4327 (1997),
Phys. Rev. Lett. 81, 4806 (1998),
Phys. Rev. Lett. 81, 5742 (1998), 
Phys. Rev. Lett. 82, 3206 (1999), 
Phys. Rev. Lett. 85, 2056 (2000).

\bibitem{lep_ff_pair}  

LEP Electroweak Working group, $\rm f \bar{\rm f}$ Subgroup, 
``Combination of the LEPII  $\rm f \bar{\rm f}$ Results'', 
LEP2FF/01-01, 
ALEPH 2001-039 PHYSIC 2001-013, 
DELPHI 2001-108 PHYS 896, 
L3 Note 2663, 
OPAL TN690, 
http://lepewwg.web.cern.ch/LEPEWWG/lep2/.



\bibitem{tev_ci}

CDF Collaboration, F.~Abe et al.,
 Phys.\ Rev.\ Lett. {\bf 79},~2198~(1997);

DO Collaboration, D.~Abbott et al.,
Phys.\ Rev.\ Lett. {\bf 82},~4769~(1999).



\bibitem{hera_lfv}

ZEUS Collab., S. Chekanov et al., Phys. Rev. D65 (2002) 092004;

ZEUS Collab., 
\textit{Search for lepton-flavor violation in $e^+p$ collisions at 
a center-of-mass-energy of 318~GeV at HERA},
contributed paper \#906 
to International Conference on High Energy Physics ``ICHEP 2002'' (July 2002). 


\bibitem{hera_fstar}

H1 Collab., C. Adloff et al., DESY-02-096, submitted to Phys. Lett. B;\\
H1 Collab., C. Adloff et al., Phys. Lett. B525 (2002) 9;


ZEUS Collaboration, J.Breitweg et al., 
Zeitschrift f. Physik C76 (1997) 4, 631-646;\\
ZEUS Collab., 
\textit{Search for Excited Fermions in ep Collisions at HERA},
Contributed paper \#912
to International Conference on High Energy Physics ``ICHEP 2002'' (July 2002), 
DESY-01-132, submitted to Physics Letters B.

\bibitem{h1_multie}
H1 Collab., \textit{Multi-Electron Production at High Transverse
 Momentum in ep Collisions at HERA},
H1prelim-02-052, contributed paper \#1019 
to International Conference on High Energy Physics ``ICHEP 2002'' (July 2002).

\bibitem{zeusee} ZEUS Collaboration,
{\it Study of multi-lepton production with the ZEUS detector at  HERA},
Contributed paper \#910 to the XXXth International Conference 
 on High Energy Physics ``ICHEP 2002'' (July 2000).

\bibitem{HERA-upgrade} W.~Bartel, E.~Gianfelice, J.~Maidment,
B.~Parker, N.~Holtkamp, E.~Lohrmann and D.~Pitzl,
{\it Prepared for Workshop on Future Physics at HERA 
(Preceded by meetings 25-26 Sep 1995 and 7-9 Feb 1996 at DESY), Hamburg, 
Germany, 30-31 May 1996}.

\bibitem{Eliana-proc} D.P. Barber, E. Gianfelice, {\it Acta Phys. Pol.} {\bf B33}, 3943 
(2002), these proceedings.

\bibitem{Jenny-proc} J. B\"ohme, {\it Acta Phys. Pol.} {\bf B33}, 3949 (2002),
these proceedings.

\bibitem{dilepton} CDF collaboration, 
Phys. Rev. D50, 2966 (1994), Phys. Rev. Lett. 73, 225 (1994), 
Phys. Rev. Lett. 74, 2626 (1995).

\bibitem{breport}  K. Anikeev et al.,
 B Physics at the Tevatron: Run II and Beyond - FERMILAB-Pub-01/197.

\bibitem{wolter} M. Wolter, {\it Acta Phys. Pol.} {\bf B33}, 2915 (2002),
these proceedings.

\bibitem{sugra} V. Barger, C.E.M.~Wagner, et al., 
Report of the SUGRA Working Group for Run II of the Tevatron -hep-ph/0003154.

\bibitem{gmsb} Ray Culbertson et al.,Low-Scale and Gauge-Mediated 
Supersymmetry Breaking at the Fermilab Tevatron Run II, hep-ph/0008070.


\bibitem{lhc_phys}
    ATLAS Collaboration, \textit{ATLAS Detector and Physics Performance
    Technical Design Report}, CERN LHCC 98-16.

\bibitem{bh} S.B. Giddings and S. Thomas, hep-ph/0106219.

\bibitem{LHCLC} G. Weiglein et al., see
http://www.cpt.dur.ac.uk/ $\tilde{}$ georg/lhclc/.

\bibitem{polarisation} G.~Moortgat-Pick and H.~M.~Steiner,
Eur.\ Phys.\ J.\ directC {\bf 6} (2001) 1
[arXiv:hep-ph/0106155]; J.~Erler {\it et al.}, hep-ph/0112070;
see also the POWER working group:
http://www.desy.de/ $\tilde{}$ gudrid/power.html.

\bibitem{TDR} 
J.~A.~Aguilar-Saavedra {\it et al.}  [ECFA/DESY LC Physics Working Group
                  Collaboration], arXiv:hep-ph/0106315.

\bibitem{Filip} P. Niezurawski, A.F. Zarnecki, M. Krawczyk,
hep-ph/0208234, to be publ. in {\it Acta Phys. Pol.} {\bf B}.


\end{thebibliography}
\end{document}